\documentclass[letterpaper,twoside,twocolumn,journal]{IEEEtran}
\usepackage[T1]{fontenc}
\usepackage[utf8]{inputenc}
\usepackage[unicode=true,
bookmarks=true,bookmarksnumbered=false,bookmarksopen=false,
breaklinks=true,pdfborder={0 0 0},pdfborderstyle={},backref=section,colorlinks=true]
{hyperref}
\pdfoutput=1
\usepackage{color}
\usepackage{array}
\usepackage{units}
\usepackage{mathtools}
\usepackage{amsmath}
\usepackage{amssymb}
\usepackage{graphicx}
\usepackage[numbers]{natbib}

\makeatletter

\pdfpageheight\paperheight
\pdfpagewidth\paperwidth

\newcommand{\noun}[1]{\textsc{#1}}
\providecommand{\tabularnewline}{\\}

\usepackage{graphicx}
\setkeys{Gin}{width=\linewidth}
\usepackage{slashbox}
\newcommand{\berkertitle}{Extensionless Adaptive Transmitter and Receiver Windowing of Beyond 5G Frames}
\usepackage{color} 
\definecolor{mygreen}{RGB}{28,172,0} 
\definecolor{mylilas}{RGB}{170,55,241}

\usepackage{newtxmath}
\usepackage{amsmath,amssymb}
\usepackage{bm}

\usepackage{breqn}
\setkeys{breqn}{labelprefix={eq:}}

\DeclareMathOperator*{\argmin}{\arg\,\min}
\DeclareMathOperator*{\argmax}{\arg\,\max}
\newcommand*{\var}[1]{\operatorname{var}\left[#1\right]}

\newcommand*{\expect}[2][]{\operatorname{E}_{#1}\left[#2\right]}

\DeclarePairedDelimiter\round{\lfloor}{\rceil}

\newcommand*{\matr}[1]{\mathbf{\bm{#1}}}
\newcommand*{\matrgr}[1]{\bm{#1}}

\newcommand*{\tran}[1]{{#1}^{\mkern-1.5mu\mathsf{T}}}
\newcommand*{\conj}[1]{{#1}^{\mkern-1.5mu \ast}}
\newcommand*{\herm}[1]{{#1}^{\mkern-1.5mu\mathsf{H}}}

\DeclareMathOperator{\vect}{vec}

\usepackage{dblfloatfix}
\usepackage{balance}
\usepackage{cite}

\usepackage{prettyref}
\newrefformat{fig}{Fig.~\ref{#1}}
\newrefformat{conj}{Conjecture~\ref{#1}}
\newrefformat{tab}{Table~\ref{#1}}
\newrefformat{sec}{Section~\ref{#1}}
\newrefformat{subsec}{Section~\ref{#1}}
\newrefformat{subsubsec}{Section~\ref{#1}}
\usepackage{algpseudocode}

\newcounter{tempEquationCounter} 
\newcounter{thisEquationNumber}

\usepackage{acronym}
\input{acros}

\usepackage{siunitx}
\DeclareSIUnit{\belmilliwatt}{Bm}
\DeclareSIUnit{\dBm}{\deci\belmilliwatt}
\DeclareSIQualifier\isotropic{i}
\DeclareSIQualifier\carrier{c}

\usepackage[caption=false,font=footnotesize]{subfig}

\newcommand\Algphase[2]{%
\vspace*{-.7\baselineskip}\Statex\hspace*{\dimexpr-\algorithmicindent-2pt\relax}\rule{\linewidth}{0.4pt}%
\Statex\hspace*{-\algorithmicindent}\textbf{{#1}:} {#2}%
\vspace*{-.5\baselineskip}\Statex\hspace*{\dimexpr-\algorithmicindent-2pt\relax}\rule{\linewidth}{0.4pt}%
}

\usepackage{cleveref}

\makeatother

\begin{document}
\title{\berkertitle}
\author{Berker Peköz, \IEEEmembership{Graduate Student Member, IEEE,} Selçuk
Köse\IEEEmembership{, Member, IEEE} and Hüseyin Arslan\IEEEmembership{, Fellow, IEEE}\thanks{Copyright (c) 2015 IEEE. Personal use of this material is permitted. However, permission to use this material for any other purposes must be obtained from the IEEE by sending a request to pubs-permissions@ieee.org.}
\thanks{This work was supported in part by the National Science Foundation under Grant 1609581.}
\thanks{Berker Pek\"{o}z is with the Department of Electrical Engineering, University of South Florida, Tampa, FL 33620 USA. (e-mail: pekoz@usf.edu)}
\thanks{Sel\c{c}uk K\"{o}se was with the Department of Electrical Engineering, University of South Florida, Tampa, FL 33620 USA. He is now with the Department of Electrical and Computer Engineering, University of Rochester, Rochester, NY 14627 USA. (e-mail: skose@ece.rochester.edu)}
\thanks{H\"{u}seyin Arslan is with the Department of Electrical Engineering, University of South Florida, Tampa, FL 33620 USA and also with the Department of Electrical and Electronics Engineering, Istanbul Medipol University, Istanbul, 34810 Turkey (e-mail: arslan@usf.edu).} }
\maketitle
\begin{abstract}
Newer cellular communication generations are planned to allow asynchronous
transmission of multiple numerologies (waveforms with different parameters)
in adjacent bands, creating unavoidable \acl{aci}. Most prior work
on windowing assume additional extensions reserved for windowing,
which does not comply with standards. Whether windowing should be
applied at the transmitter or the receiver was not questioned. In
this work, we propose two independent algorithms that are implemented
at the transmitter and receiver, respectively. These algorithms estimate
the transmitter and receiver windowing duration of each \ac{re} with
an aim to improve fair proportional network throughput. While doing
so, we solely utilize the available extension that was defined in
the standard and present standard-compliant algorithms that also do
not require any modifications on the counterparts or control signaling.
Furthermore, computationally efficient techniques to apply per-\ac{re}
transmitter and receiver windowing to signals synthesized and analyzed
using conventional \acl{cp}-\acl{ofdm} are derived and their computational
complexities are analyzed. The spectrotemporal relations between
optimum window durations at either side, as well as functions of the
excess \aclp{snr}, the subcarrier spacings and the throughput gains
provided over previous similar techniques are numerically verified.\acresetall
\end{abstract}


\IEEEpeerreviewmaketitle{}

\section{Introduction}

\ac{3gpp} designed 4G-\ac{lte} to deliver broadband services to
masses\citep{astely2009ltethe}. The design was successful in doing
what it promised, but the one-size-fits-all approach resulted in certain
engineering trade-offs. This broadband experience was possible at
a certain reliability not allowing \ac{urllc} operations, is not
the most power-efficient design and is only possible below $\SI[per-mode=symbol]{120}{\kilo\meter\per\hour}$
mobility\citep{andrews2014whatwill}. 5G \ac{nr} physical layer was
designed to utilize the \ac{ofdm} waveform\citep{chang1966synthesis}
with different parameters, called numerologies, allowing prioritization
of certain aspects in different applications and made the \ac{embb}
experience possible in a wider range of scenarios\citep{chen2014therequirements}.
For example, while low power \ac{iot} devices are assigned smaller
subcarrier spacings to conserve battery, vehicular communications
are operating with higher subcarrier spacings and shorter symbol durations
to keep the communication reliable in high Doppler spreads caused
by higher speeds.

\begin{figure}
\subfloat[\label{fig:Fig1a}]{\includegraphics{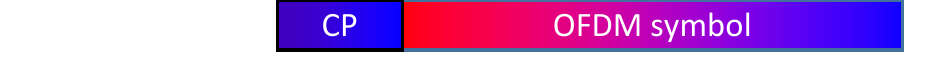}

}

\subfloat[\label{fig:Fig1b}]{\includegraphics{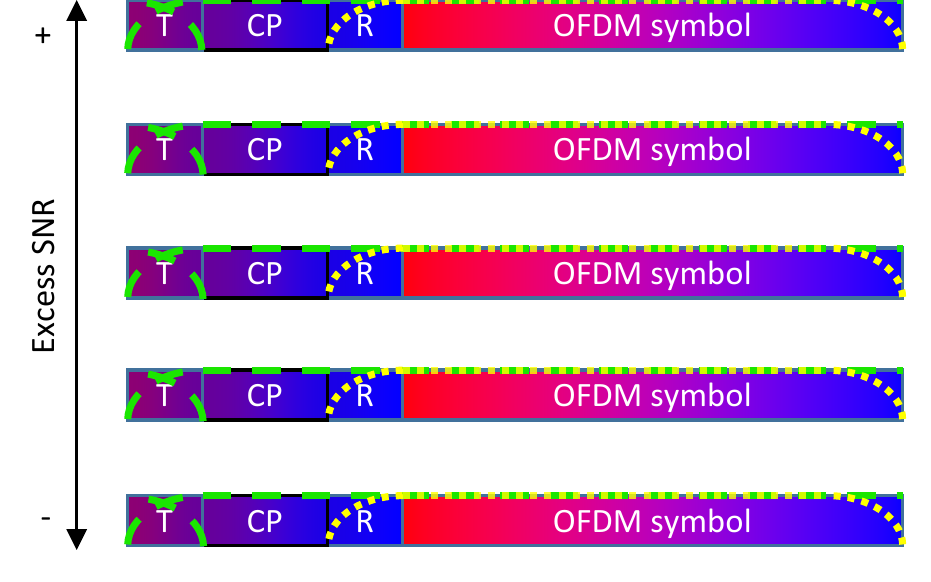}

}

\subfloat[\label{fig:Fig1c}]{\includegraphics{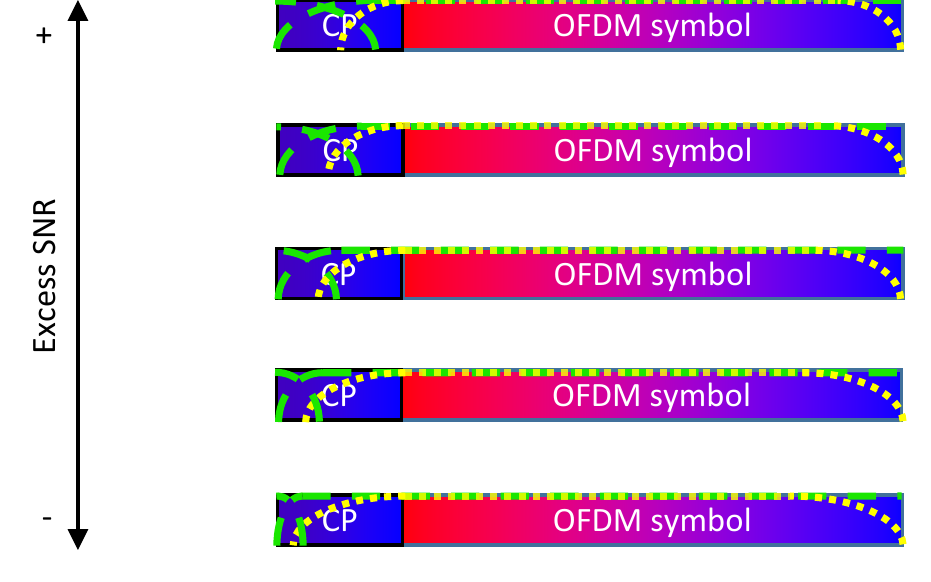}

}

\caption{Visual demonstrations of temporal (a) standard symbol structure, (b)
structure used in previous windowing literature and (c) the adaptive
\ac{cp} concept presented in this work. The rectangles are alotted
times for the actual \ac{ofdm} symbol, \ac{cp}, and further cyclic
extensions for ``T''ransmitter and ``R''eceiver windowing, while
the green dash and yellow round dot overlays demonstrate transmitter
and receiver windowing of the underlying area, respectively.\label{fig:A-visual-demonstration}}
\end{figure}

This shift in paradigm brought with it a problem deliberately avoided
by the uniform design. Regardless of the domain \ac{ma} was performed,
the use of a unified orthogonal waveform in the point-to-multipoint
\ac{dl} avoided the inter-\ac{ue} interference problem experienced
in the multipoint-to-point \ac{ul} in all preceding generations of
cellular communications. However, by allowing coexistence of different
\ac{ofdm} numerologies in adjacent bands, \ac{aci} between \acp{ue}
sharing these bands arises in the \ac{dl}\citep{ankarali2017flexible}.
In the \ac{ul}, although orthogonal waveforms were used in principle,
power differences and timing and frequency offsets across \acp{ue}
caused interference. Although they came at certain costs, strict timing
and frequency synchronization across \acp{ue}\citep{3gpp.38.214}
and power control\citep{lupas1990nearfar} have been historically
used to mitigate the interference in the \ac{ul}. Unfortunately,
with the use of different numerologies, these remedies are not a solution
to the problem and \ac{ini} is inevitable\citep{Zhang2018} even
in the \ac{dl}. \ac{3gpp} acknowledges this problem and gives manufacturers
the freedom to implement any solution they choose as long as they
respect the standard frame structure\citep{3gpp.38.201} seen in \prettyref{fig:Fig1a}.

Windowing of \ac{ofdm} signals is a well-studied interference management
technique that has garnered attention due to its low computational
complexity. Windowing can be performed independently at the transmitter
to reduce \ac{oob} emission\citep{gudmundson1996adjacent}, or at
the receiver to reduce interference caused by communication taking
place in adjacent channels, commonly referred to as \ac{aci} \citep{muschallik1996improving}.
Most recently in \citep{guvenkaya2015awindowing}, utilizing different
window functions for each subcarrier at the transmitter and receiver
is proposed and the window functions for each subcarrier that maximizes
the spectral localization within the \ac{ue}'s resources and interference
rejection are derived.

In \citep{guvenkaya2015awindowing} and most of the preceding literature
focusing on windowing, windowing was performed by extending the symbols
by an amount which was arbitrarily chosen without explanation, in
addition to standard \ac{cp} duration seen in \prettyref{fig:Fig1b},
and the focus was on deriving window functions optimized according
to maximize standard performance metrics. These extensions reduce
the symbol rate and change the frame structure defined in the standard,
thus creating nonstandard signals that are not orthogonal to the symbols
that aims to share the same numerology\citep{ankarali2017flexible}.
As mentioned above, this is not acceptable in the current cellular
communication standards\citep{3gpp.38.201}. Furthermore, extending
the symbol duration relentlessly causes the symbol duration to exceed
the coherence time of the channel, which is a critical problem for
high-speed vehicular communications\citep{ankarali2017flexible}.
In \citep{Sahin2011}, the authors attempted to improve spectral efficiency
of windowed \ac{ofdm} systems by not applying windowing to the \acp{re}
of inner subcarriers assigned to \acp{ue} experiencing long delay
spreads and applying windowing on the edge subcarriers using the excess
\ac{cp} assigned to \acp{ue} experiencing short delay spreads. While
effective, this scheme is only applicable if all \acp{ue} utilize
the same numerology. The first standard compliant windowing scheme
was proposed in \citep{pekoz2017adaptive}, in which the authors derived
the receiver windowing durations that optimize reception of each subcarrier
in the case which \ac{isi} and \ac{aci} occur simultaneously and
pulse shapes of transmitters operating in adjacent bands cannot be
controlled, in the absence of any extension designated for windowing.
Whether it is more beneficial to window a duration at the transmitter
or receiver was not discussed in the literature.

This work aims to extend \citep{pekoz2017adaptive} by evaluating
how network capacity can be further improved if the pulse shapes of
the transmitted waveforms can also be designed while conserving the
standard frame structure, that is, not adding any additional extensions
other than \ac{cp} and using only the present \ac{cp} for windowing.
In this work, we propose two independent algorithms that aim to determine
the amount of windowing that should be applied at either side to maximize
fair proportional network capacity. Unlike \citep{pekoz2017adaptive}
in which receiver windowing duration calculations required \acp{cir}
knowledge, the proposed receiver windowing duration calculation algorithm
in this work is solely uses statistics derived from the received signals.
This significantly reduces the complexity and eases implementation,
and makes the algorithm completely practical as no information is
needed. The proposed transmitter windowing duration calculation algorithm
aims to maximize the network spectral efficiency by assigning high
transmit window durations only to \acp{re} with excess \ac{sinr}
that can withstand the \ac{isi} caused by windowing. This reduces
the \ac{aci} in the system with minimum impact to the \acp{re} applying
windowing. Neither algorithm requires any control data transfer to
other parties of the communication or changes to the other nodes at
any point. The proposed utilization of the standard symbol structure
as a function of excess \ac{snr} is shown in \prettyref{fig:Fig1c}.
Numerical results confirm that fair proportional network spectral
efficiency can be increased greatly without disrupting the standard
frame structure by utilizing \ac{cp} adaptively, and determining
transmitter windowing durations using excess \ac{sinr} of \acp{re}
and data-aided receiver windowing duration determination are an effective
metrics.
Our contributions in this work are as follows:
\begin{itemize}
\item Computationally efficient per-\ac{re} transmitter and receiver windowing
of signals synthesized and analyzed using conventional \ac{cp}-\ac{ofdm}
are derived.
\item A computationally efficient per-\ac{re} transmitter window duration
estimation algorithm for \acp{gnb} that maximizes the fair proportional
network throughput based on \acp{ue} channel conditions and does
not require any information transfer to or modification at \acp{ue}
is presented.
\item A computationally efficient per-\ac{re} receiver window duration
estimation algorithm for \acp{gnb} and \acp{ue} that maximizes the
capacity of each \ac{re} and does not require any information transfer
from or modification at the transmitter is presented.
\item The computational complexities of the aforementioned algorithms are
derived.
\item The algorithms are numerically analyzed in terms of \ac{oob}-emission
reduction, throughput improvements, relation of window duration estimates
with excess \ac{snr}, spectrotemporal correlation and accuracy of
window duration estimates.
\end{itemize}
\textit{Notation}: $\tran{\left(\cdot\right)}$, $\conj{\left(\cdot\right)}$
and $\herm{\left(\cdot\right)}$ denote the transpose, conjugate and
Hermitian operations, $\matr{A}\left[a,b\right]$ is the element in
the $a$th row and $b$th column of matrix $\matr{A}$, $\matr{A}\left[a,:\right]$
and $\matr{A}\left[:,b\right]$ are each row and column vectors containing
the $a$th row and $b$th column of matrix $\matr{A}$, respectively,
$\vect{\left(\matr{A}\right)}=\tran{\begin{bmatrix}\tran{\matr{A}\left[:,1\right]} & \tran{\matr{A}\left[:,2\right]} & \dots\end{bmatrix}}$
is the vectorization operator, $\matr{A}\odot\matr{B}$ and $\matr{A}\oslash\matr{B}$
correspond to Hadamard multiplication and division of matrices $\matr{A}$
and $\matr{B}$ and $\matr{A}$ by $\matr{B}$,  $\matr{0}_{A\times B}$
denotes matrices of zeros with $A$ rows and $B$ columns, $\mathcal{CN}\left(\mu,\sigma^{2}\right)$
represents complex Gaussian random processes with mean $\mu$ and
variance $\sigma^{2}$, $\round{\matr{X}}$ correspond to rounding
all elements of $\matr{X}$ to the nearest integer, $\#\mathbb{S}$
denotes the cardinality of set $\mathbb{S}$, $\expect[x]{\matr{y}}$
is the expected value of random vector $\matr{y}$ with respect to
variable $x$, and $\jmath=\sqrt{-1}$.

\section{System Model\label{sec:System-Model}}

In this work, we assume that there is a node, referred to as the \ac{gnb},
that conveys information to all other nodes in the system and all
other nodes aim to convey information solely to the \ac{gnb} during
processes referred to as \ac{dl} and \ac{ul}, respectively. There
are $U$ nodes other than the \ac{gnb}, hereinafter referred to as
\acp{ue}, sharing a total bandwidth $B$ to communicate with the
\ac{gnb} using \ac{ofdm}. Each \ac{ue} $u$ samples this whole
band band using an $N_{u}$-point \ac{fft}, such that the frequency
spacing between the points at the \ac{fft} output becomes $\Delta f_{u}=B/N_{u}$.
The quantity $\Delta f_{u}$ is referred to as the subcarrier spacing
of \ac{ue} $u$. Bi-directional communication takes place in a \ac{tdd}
fashion and \ac{fdma} is used for multiple accessing; \acp{ue} solely
receive and do not transmit during \ac{gnb}'s transmission, i.e.
\ac{dl}, whereas all \acp{ue} transmit simultaneously in adjacent
but non-overlapping frequency bands in the \ac{ul}. \ac{ul} is assumed
to take place before \ac{dl} and is crucial to the work, but we focus
on modeling the details of the \ac{dl} necessary for the proposed
methods for sake of brevity, while necessary details regarding \ac{ul}
are provided in numerical verification. The data of each \ac{ue}
$u$ is conveyed in $M_{u}$ consecutive subcarriers of $L_{u}$ consecutive
\ac{ofdm} symbols, with contiguous indices $\left\{ M_{u,1},\dots,M_{u,M_{u}}\right\} $
out of the possible $N_{u}$, while the remaining subcarriers are
left empty for use by other \acp{ue}.  Although the algorithms presented
and performed analysis are directly compatible with \ac{ofdma}, for
the sake of simplifying the notation throughout this work, we assume
pure \ac{fdma}, that is, $L_{u_{1}}N_{u_{1}}=L_{u_{2}}N_{u_{2}}\,,\forall u_{1},u_{2}\in\mathbb{N}_{\leq U}^{*}$.

Symbols known by receiving nodes, commonly referred to as pilot or
\ac{dmrs}, are transmitted in some \acp{re} for time synchronization
and channel estimation purposes in both \ac{ul} and \ac{dl}. The
\ac{dmrs} transmitted to \ac{ue} $u$ are contained in the sparse
matrix $\matr{P}_{u}\in\mathbb{C}^{M_{u}\times L_{u}}$. The \ac{sc}
data symbols transmitted to \ac{ue} $u$ are contained in $\matr{D}_{u}\in\mathbb{C}^{M_{u}\times L_{u}}$,
of which nonzero  elements do not overlap with that of $\matr{P}_{u}$.

A \ac{cp} of length $K_{u}$ samples is appended to the each time
domain \ac{ofdm} symbol to mitigate multipath propagation and prevent
\ac{isi}, where $K_{u}/L_{u}$ equals to the same constant for all
\acp{ue} of the network and is referred to as the \ac{cp} rate.
The \ac{ofdm} symbol samples, preceded their respective \ac{cp}
samples to be broadcasted to all users can be obtained as 
\begin{equation}
\breve{\matr{x}}=\sum_{u=1}^{U}\vect{\left(\begin{bmatrix}\begin{array}{cc}
\matr{0}_{K_{u}\times\left(N_{u}-K_{u}\right)} & \matr{I}_{K_{u}}\end{array}\\
\matr{I}_{N_{u}}
\end{bmatrix}\matr{F}_{N_{u}}\matr{Q}_{u}\left(\matr{P}_{u}+\matr{D}_{u}\right)\right)},\label{eq:basetxsymgen}
\end{equation}
where $\breve{\matr{x}}\in\mathbb{C}^{\left(N_{u}+K_{u}\right)L_{u}\times1,\,\forall u}$
is the basic baseband sample sequence, $\matr{Q}_{u}\in\mathbb{R}^{N_{u}\times M_{u}}$
is the resource mapping matrix of $u$th \ac{ue} that maps the data
elements to the scheduled resources, and $\matr{F}_{N_{u}}\in\mathbb{C}^{N_{u}\times N_{u}}$
is the normalized $N_{u}$-point \ac{fft} matrix. Some \ac{cp} samples
may also be used for transmitter windowing to limit the \ac{oob}
emission as described in \citep{guvenkaya2015awindowing,bala_shaping_2013}.
Different transmitter window durations may be utilized for each \ac{re}
to be transmitted to each \ac{ue}. The transmitter window durations
associated with $u$th \ac{ue}'s \acp{re} are given in $\matr{T}_{u}\in\mathbb{N}_{\leq K_{u}}^{M_{u}\times L_{u}}$
and  calculated according to \prettyref{subsec:Estimation-of-Optimum}.
Let $\matr{x}\in\mathbb{C}^{\left(\left(K_{u}+N_{u}\right)L_{u}\right)\times1}$
denote the per-\ac{re} transmitter windowed baseband sample sequence,
calculated computationally efficiently as described in \prettyref{subsec:Estimation-of-Optimum}.

 The waveform is then transmitted over the multiple access multipath
channel. The complex channel gain of the cluster that arrives at the
$u$th \ac{ue} at the $t$th sample after a delay of $\tau$ samples
is denoted by the complex coefficient $h_{u,\tau,t}$. We assume that
these channel gains are normalized such that $\mathbb{E}_{t}\left[\sum_{\tau=0}^{t-\Delta_{t,u}-1}\left|h_{u,\tau,t}\right|^{2}\right]=1$
and that they vary at each sample instant where the mobility of each
\ac{ue} is independent of all others. Then, the $t$th sample received
at $u$th \ac{ue} is written as 
\begin{equation}
\matr{y}_{u}\left[t\right]=\tilde{n}+\sqrt{\gamma_{u}}\sum_{\tau=0}^{t-\Delta_{t,u}-1}h_{u,\tau,t}\matr{x}\left[t-\Delta_{t,u}-\tau\right],\,t\in\mathbb{N}^{*},
\end{equation}
where $\matr{x}\left[t\right]\coloneqq0,\,\forall t\in\mathbb{N}_{>\left(K_{u}+N_{u}\right)L_{u}}\cup\mathbb{Z}^{-}$,
$\tilde{n}\sim\mathcal{CN}\left(0,1\right)$ is the background \ac{awgn},
$\gamma_{u}$ is the overall \ac{snr} of $u$th \ac{ue} and $\Delta_{t,u}$
is the propagation delay for $u$th \ac{ue} in number of samples.
Each \ac{ue} then synchronizes to their signal by correlating the
received samples with samples generated only using their $\matr{P}_{u}$
and estimates $\hat{\Delta_{t,u}}$\citep{wang2003achannel}. The
samples estimated to contain $u$th \ac{ue}'s $l$th \ac{ofdm} symbol
and its corresponding \ac{cp} is denoted by vector $\matr{y}_{l,u}\in\mathbb{C}^{\left(K_{u}+N_{u}\right)\times1}$,
where $\matr{y}_{l,u}\left[s\right]=\matr{y}_{u}\left[\left(l-1\right)\left(N_{u}+K_{u}\right)+\hat{\Delta_{t,u}}+s\right],\,s\in\mathbb{N}_{\leq\left(K_{u}+N_{u}\right)}^{*}$.
Before the receiver windowing operations are performed, $u$th \ac{ue}
first performs regular \ac{ofdm} reception and calculates the received
\ac{sc} symbols from the \ac{ofdm} symbol samples as
\begin{equation}
\matr{Y}_{u}\left[:,l,0\right]=\tran{\matr{Q}_{u}}\matr{F}_{N_{u}}\begin{bmatrix}\matr{0}_{N_{u}\times K_{u}} & \matr{I}_{N_{u}}\end{bmatrix}\matr{y}_{l,u}^{\text{}},\label{eq:rxscsyms}
\end{equation}
where the first plane of $\matr{Y}_{u}\in\mathbb{C}^{M_{u}\times L_{u}\times\left(K_{u}+1\right)}$
are the received base \ac{sc} symbols. Each \ac{ue} uses a different
receiver window duration to receive each \ac{re}. The receiver window
durations associated with $u$th \ac{ue}'s \acp{re} are given in
$\matr{R}_{u}\in\mathbb{N}_{\leq K_{u}}^{M_{u}\times L_{u}}$ and
are calculated according to \prettyref{subsec:Optimum-Receiver-Window},
wherein also the calculation of the receiver windowed \ac{sc} symbols
$\hat{\matr{Y}}_{u}\in\mathbb{C}^{M_{u}\times L_{u}}$ are demonstrated.
\Ac{cfr} coefficients at \ac{dmrs} locations are first estimated
as 
\begin{equation}
\breve{\matr{H}}_{u}\left[m,l\right]=\hat{\matr{Y}}_{u}\left[m,l\right]\oslash\matr{P}_{u}\left[m,l\right]
\end{equation}
using nonzero elements of $\mathbf{P}_{u}$. Then, a transform domain
channel estimator \citep[(33)]{ozdemir2007channel} is applied and
estimated \acp{cir} are reduced to their first $K_{u}$ coefficients.
The CIR coefficients of non-pilot carrying symbols are interpolated
and extrapolated\citep{oien2004impactof}, and all \ac{cfr} coefficient
estimates $\hat{\matr{H}}_{u}$ are obtained\citep[(33)]{ozdemir2007channel}.
Finally, data symbols are equalized as described in \citep{brady1970adaptive}
for nonzero elements of $\mathbf{D}_{u}$ and the received symbols
are estimated as 
\begin{equation}
\hat{\mathbf{D}}_{u}=\frac{\hat{\matr{Y}}_{u}\odot\conj{\text{\ensuremath{\hat{\matr{H}}}}}_{u}}{\hat{\matrgr{\sigma}}_{\text{n},u}^{2}+\text{\ensuremath{\hat{\matr{H}}_{u}}}\odot\conj{\text{\ensuremath{\hat{\matr{H}}}}}_{u}},
\end{equation}
where $\hat{\matrgr{\sigma}}_{\text{n},u}^{2}\in\mathbb{R}^{M_{u}\times L_{u}}$
is the variance estimated by $u$th \ac{ue} for noise, various interference
sources and other disruptions.

\section{Proposed Method\label{sec:Proposed-Method}}

The idea proposed in this work involves determination of $\matr{T}_{u}$
and $\matr{R}_{u}$ for all \acp{ue} that maximizes the fair-proportional
network capacity. Because these concepts are implemented independently,
they are discussed separately.

\subsection{Estimation of Optimum Transmitter Window Durations\label{subsec:Estimation-of-Optimum}}

This subsection first discusses efficient differential calculation
of per-\ac{re} transmitter windowed samples to prove the optimum
transmitter window durations calculations feasible. The optimization
metric, fair proportional network capacity, is then defined. An algorithm
to effectively maximize the fair proportional network capacity is
provided. Finally, the computational complexity of the provided algorithm
is calculated and discussed.

\subsubsection{Computationally Efficient Conversion of Conventional \ac{cp}-\ac{ofdm}
Samples to Per-\ac{re} Transmitter Windowed \ac{ofdm} Samples}

\begin{figure}
\includegraphics{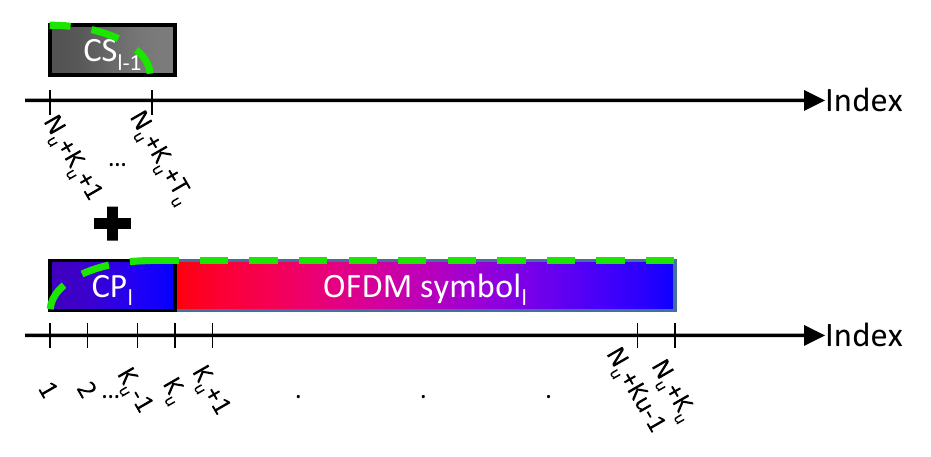}

\caption{Indexing of $\matr{t}$ within a demonstration of how transmitter
windowed samples are generated by overlapping scaled \ac{cp} of current
and \ac{cs} of preceding \ac{ofdm} symbols of which indices are
given in the subscripts. \label{fig:Fig2}}
\end{figure}

The transmit pulse shape of the $m$th subcarrier of $l$th \ac{ofdm}
symbol to be transmitted to \ac{ue} $u$ in accordance with $\matr{T}_{u}\left[m,l\right]$
is contained in the vector $\matr{t}_{m,l,u}\in\mathbb{R}^{\left(K_{u}+N_{u}+T_{u}\right)\times1}$
of which indexing is demonstrated in \prettyref{fig:Fig2} and is
calculated per \citep{guvenkaya2015awindowing} to contain the energy
of that subcarrier within the band assigned to the \ac{ue}. Investigating
\prettyref{eq:basetxsymgen}, if no transmitter windowing is applied,
i.e. the generation of a regular \ac{cp}-\ac{ofdm} sample sequence
$\breve{\matr{x}}$, the contribution from the symbol in the $m$th
subcarrier of the $l$th \ac{ofdm} symbol of $u$th user to the $k\leq K_{u}$th
sample of that \ac{ofdm} symbol is $\exp\left(-\jmath2\pi M_{u,m}\left(k-K_{U}-1\right)/N_{u}\right)\left(\matr{D}_{u}\left[m,l\right]+\matr{P}_{u}\left[m,l\right]\right)/\sqrt{N_{u}}$.
If transmitter windowing is applied to the \ac{re} in interest, the
contribution at $k\leq\matr{T}_{u}\left[m,l\right]$th sample would
instead become
\begin{dseries}\begin{math}
\left(\matr{t}_{m,l,u}\left[k\right]\exp\left(-\jmath\tfrac{2\pi M_{u,m}\left(k-1-K_{U}\right)}{N_{u}}\right)\left(\matr{D}_{u}\left[m,l\right]+\matr{P}_{u}\left[m,l\right]\right)+\matr{t}_{m,l,u}\left[k+N_{u}+K_{u}\right]\exp\left(-\jmath\tfrac{2\pi M_{u,m}\left(k-1\right)}{N_{u}}\right)\left(\matr{D}_{u}\left[m,l-1\right]+\matr{P}_{u}\left[m,l-1\right]\right)\right)/\sqrt{N_{u}}\end{math}\begin{math}
=\frac{\exp\left(-\jmath\tfrac{2\pi M_{u,m}\left(k-1\right)}{N_{u}}\right)}{\sqrt{N_{u}}}\left(\matr{t}_{m,l,u}\left[k\right]\exp\left(\jmath\tfrac{2\pi M_{u,m}K_{U}}{N_{u}}\right)\left(\matr{D}_{u}\left[m,l\right]+\matr{P}_{u}\left[m,l\right]\right)+\matr{t}_{m,l,u}\left[k+N_{u}+K_{u}\right]\left(\matr{D}_{u}\left[m,l-1\right]+\matr{P}_{u}\left[m,l-1\right]\right)\right).\end{math}
\end{dseries}Noting that $\matr{t}_{m,l,u}\left[k\right]\coloneqq1-\matr{t}_{m,l,u}\left[k+N_{u}+K_{u}\right],\,\forall k\in\mathbb{Z}_{\leq N_{u}+K_{u}}^{+}$\citep{guvenkaya2015awindowing,bala_shaping_2013},
the $k\leq T_{u}\left[m,l\right]$th sample of $u$th user's $l$th
\ac{ofdm} symbol's per-\ac{re} transmitter windowed $m$th subcarrier
can be converted from that generated using a conventional \ac{cp}-\ac{ofdm}
procedure as\begin{dmath}
\matr{x}\left[\left(l-1\right)\left(N_{u}+K_{u}\right)+k\right]=\breve{\matr{x}}\left[\left(l-1\right)\left(N_{u}+K_{u}\right)+k\right]+\tfrac{\matr{t}_{m,l,u}\left[k+N_{u}+K_{u}\right]\exp\left(-\jmath\tfrac{2\pi M_{u,m}\left(k-1\right)}{N_{u}}\right)}{\sqrt{N_{u}}}\left(\left(\matr{D}_{u}\left[m,l-1\right]+\matr{P}_{u}\left[m,l-1\right]\right)-\exp\left(\jmath\tfrac{2\pi M_{u,m}K_{U}}{N_{u}}\right)\left(\matr{D}_{u}\left[m,l\right]+\matr{P}_{u}\left[m,l\right]\right)\right).\label{eq:txwinconvertsamp}
\end{dmath}$\matr{x}$ can be obtained by converting all $T_{u}\left[m,l\right]$
samples of $\breve{\matr{x}}$ to per-\ac{re} transmitter windowed
samples, and this is implied in all further references to \prettyref{eq:txwinconvertsamp}.

\subsubsection{Estimation of Fair Proportional Network Capacity}

In order to estimate the optimum transmitter window durations, the
\ac{gnb} first estimates the \ac{sinr} and corresponding capacity
for each \ac{re} of each user prior to transmission, calculates the
fair proportional network capacity, and tries to increase it iteratively.
The samples to be received at the $u$th \ac{ue} are first estimated
as
\begin{equation}
\hat{\matr{y}}_{u}\left[t\right]=\sum_{\tau=0}^{t-1}\hat{h}_{u,\tau,t}\matr{x}\left[t-\tau\right],\,t\in\mathbb{N}_{\leq\left(K_{u}+N_{u}\right)L_{u}}^{*},\label{eq:estimatedreceivedsamples}
\end{equation}
where $\hat{h}_{u,\tau,t}$ are the \ac{cir} coefficient predictions\citep{oien2004impactof}
at the \ac{gnb} prior to transmission\footnote{$\hat{h}_{u,\tau,t}$ inherits $\sqrt{\gamma_{u}}$ in the channel
estimation phase.}. The samples are regrouped accordingly to $L_{u}$ groups of $K_{u}+N_{u}$
samples each and receiver processed as described in \prettyref{sec:System-Model},
that is, \acp{cp} are removed from all symbols, \acp{fft} are applied,
and receiver windowing is performed if the \ac{gnb} is aware that
the receiver in current interest does so. Results for various cases
of receiver windowing are provided in \prettyref{sec:Numerical-Verification},
but for the sake of brevity, we assume that the \ac{gnb} assumes
none of the \acp{ue} perform receiver windowing in the remainder
of this subsection. The \ac{gnb} estimate at the \ac{fft} output,
$\tilde{\matr{Y}}_{u}\in\mathbb{C}^{M_{u}\times L_{u}}$, is formulated
as
\begin{equation}
\tilde{\matr{Y}}_{u}\left[m,l\right]=\tilde{\matr{H}}_{u}\left[m,l\right]\left(\mathbf{D}_{u}\left[m,l\right]+\tilde{\mathbf{D}}_{u}\left[m,l\right]\right),\label{eq:estimatedreceivedsymbols}
\end{equation}
where $\tilde{\matr{H}}_{u}\left[m,l\right]$ is the \ac{cfr} coefficient
prediction of the $m$th subcarrier of the $l$th \ac{ofdm} symbol
of $u$th user, the first term inside parentheses is due to the data
itself, and the second term inside parentheses shown as $\tilde{\mathbf{D}}_{u}$
is the cumulative \ac{aci}, \ac{ici} and \ac{isi}. Note that since
the source samples for all these effects are summed with that of data
at the \ac{gnb} and are passed through the same channel, this cumulative
disruption is also scaled with the same channel gain. Accordingly,
the number of bits that can be conveyed in the actual noisy transmission
channel using the data carrying $m$th subcarrier of the $l$th \ac{ofdm}
symbol of $u$th \ac{ue} is \citep{Shannon1948}
\begin{align}
\breve{\matrgr{\eta}}_{u}\left[m,l\right] & =\log_{2}\left(1+\frac{\left|\tilde{\matr{H}}_{u}\left[m,l\right]\right|^{2}}{1+\left|\tilde{\matr{H}}_{u}\left[m,l\right]\right|^{2}\left|\tilde{\mathbf{D}}_{u}\left[m,l\right]\right|^{2}}\right)\nonumber \\
 & =\log_{2}\left(1+\frac{\left|\tilde{\matr{H}}_{u}\left[m,l\right]\right|^{2}}{1+\left|\tilde{\matr{Y}}_{u}\left[m,l\right]-\tilde{\matr{H}}_{u}\left[m,l\right]\mathbf{D}_{u}\left[m,l\right]\right|^{2}}\right).\label{eq:uncappedcap}
\end{align}
If the \ac{re} under investigation is scheduled to use a certain
\ac{mcs} to carry $\matr{b}_{u}\left[m,l\right]$ bits, \prettyref{eq:uncappedcap}
is in fact capped as 
\begin{equation}
\matrgr{\eta}_{u}\left[m,l\right]=\min\left(\matr{b}_{u}\left[m,l\right],\breve{\matrgr{\eta}}_{u}\left[m,l\right]\right).\label{eq:CappedCapSngleSym}
\end{equation}
The mean number of bits conveyable to $u$th \ac{ue} per \ac{re}
is 
\begin{align}
\bar{\eta}_{u} & =\expect[m]{\expect[l]{\matrgr{\eta}_{u}\left[m,l\right]}},\label{eq:meanuecap}
\end{align}
and we define the fair proportional network capacity as the geometric
mean of the mean capacities of all \acp{ue} in the network
\begin{equation}
\eta=\sqrt[U]{\prod_{u=1}^{U}\bar{\eta}_{u}}.\label{eq:netcap}
\end{equation}

\subsubsection{Optimum Transmitter Window Duration Estimation Algorithm\label{subsec:Transmitter-Algorithm}}

Given the discrete nature of possible window durations in digital
pulse shaping and the lack of a relation between window duration and
amount of interference incurred on a victim subcarrier for optimum
window functions used in this work\citep{guvenkaya2015awindowing}
for the time-varying multipath multiple access channel, an analytical
solution to this multivariate integer optimization problem with such
a nonlinear utility function is not obvious at the time of writing.
The choice of transmitter window duration of any \ac{re} must balance
the \ac{sinr} degradation to the \acp{re} caused by induced \ac{isi},
and the \ac{sinr} improvement to all other \acp{re}, particularly
those of other \acp{ue}.  The transmitter window duration affects
the whole network, hence, must be calculated keeping the whole network
in mind, meaning \prettyref{eq:netcap} must be calculated and optimized
at the \ac{gnb} prior to transmission. However, explicitly calculating
\cref{eq:txwinconvertsamp,eq:estimatedreceivedsamples,eq:estimatedreceivedsymbols,eq:uncappedcap,eq:CappedCapSngleSym,eq:meanuecap,eq:netcap}
every time for each \ac{re} is computationally exhaustive. The aforementioned
equations are provided to provide the necessary understanding, but
the following equations will be used in the computationally efficient
estimation of optimum transmitter window durations. Consider that
we wish to test whether setting the transmitter window duration of
the \ac{re} in the $\dot{m}$th subcarrier of the $\dot{l}$th \ac{ofdm}
symbol of $\dot{u}$th user to $\matr{T}_{\dot{u}}\left[\dot{m},\dot{l}\right]$
improves the fair proportional network capacity or not. Assume the
transmitter windowed samples are calculated per \prettyref{eq:txwinconvertsamp}.
To keep expressions clear, let us refer to the difference in the $k$th
\ac{cp} sample in interest per \prettyref{eq:txwinconvertsamp} as
\begin{equation}
\dot{x}_{k}\coloneqq\matr{x}\left[\left(\dot{l}-1\right)\left(N_{\dot{u}}+K_{\dot{u}}\right)+k\right]-\breve{\matr{x}}\left[\left(\dot{l}-1\right)\left(N_{\dot{u}}+K_{\dot{u}}\right)+k\right]
\end{equation}
\begin{dmath}
=\tfrac{\matr{t}_{\dot{m},\dot{l},\dot{u}}\left[k+N_{\dot{u}}+K_{\dot{u}}\right]\exp\left(-\jmath\tfrac{2\pi M_{\dot{u},\dot{m}}\left(k-1\right)}{N_{\dot{u}}}\right)}{\sqrt{N_{\dot{u}}}}\left(\left(\matr{D}_{\dot{u}}\left[\dot{m},\dot{l}-1\right]+\matr{P}_{\dot{u}}\left[\dot{m},\dot{l}-1\right]\right)-\exp\left(\jmath\tfrac{2\pi M_{\dot{u},\dot{m}}K_{\dot{U}}}{N_{\dot{u}}}\right)\left(\matr{D}_{\dot{u}}\left[\dot{m},\dot{l}\right]+\matr{P}_{\dot{u}}\left[\dot{m},\dot{l}\right]\right)\right).\label{eq:txwinsampdiff}
\end{dmath}

The next step is to regenerate \prettyref{eq:estimatedreceivedsymbols}
for all \acp{ue}. However, as the number of changed symbols is limited,
whole sample sequences do not need regeneration, but only the received
samples that are affected by the changed samples, and fall into a
valid receiver window must be recalculated. For example, assuming
a conventional rectangular receiver window is utilized at the receivers,
which will be assumed in the rest of this section, any changes to
\ac{cp} samples will be discarded as they fall outside the receiver
window, hence need not be calculated. In this case, first $\matr{T}_{\dot{u}}\left[\dot{m},\dot{l}\right]$
modified samples that the channel would leak into the symbol duration
must be calculated for each \ac{ue}, and the $k$th sample (per indexing
of \prettyref{fig:Fig2}) of the transmitter windowed received sample
sequence $\widehat{\matr{y}_{u}}$ can be written as\begin{dgroup}
\begin{dmath}
\widehat{\matr{y}_{u}}\left[\left(\dot{l}-1\right)\left(N_{\dot{u}}+K_{\dot{u}}\right)+k\right]=\sum_{\tau=0}^{K_{u}}\hat{h}_{u,\tau,\left(\dot{l}-1\right)\left(N_{\dot{u}}+K_{\dot{u}}\right)+k}\matr{x}\left[\left(\dot{l}-1\right)\left(N_{\dot{u}}+K_{\dot{u}}\right)+k-\tau\right]
\end{dmath}
\begin{dmath}
=\sum_{\tau=0}^{K_{u}}\hat{h}_{u,\tau,\left(\dot{l}-1\right)\left(N_{\dot{u}}+K_{\dot{u}}\right)+k}\left(\breve{\matr{x}}\left[\left(\dot{l}-1\right)\left(N_{\dot{u}}+K_{\dot{u}}\right)+k-\tau\right]+\dot{x}_{k-\tau}\right)
\end{dmath}
\begin{dmath}
=\hat{\matr{y}}_{u}\left[\left(\dot{l}-1\right)\left(N_{\dot{u}}+K_{\dot{u}}\right)+k\right]+\sum_{\tau=k-\matr{T}_{\dot{u}}\left[\dot{m},\dot{l}\right]}^{K_{u}}\hat{h}_{u,\tau,\left(\dot{l}-1\right)\left(N_{\dot{u}}+K_{\dot{u}}\right)+k}\dot{x}_{k-\tau}.
\end{dmath}
\end{dgroup}Let us similarly refer to the difference in the $k$th relevant (belonging
to the \ac{ofdm} symbol affected by the windowing operation) sample
to be received by the $u$th \ac{ue} as 
\begin{align}
\dot{y}_{u,k} & =\widehat{\matr{y}_{u}}\left[\left(\dot{l}-1\right)\left(N_{\dot{u}}+K_{\dot{u}}\right)+k\right]-\hat{\matr{y}}_{u}\left[\left(\dot{l}-1\right)\left(N_{\dot{u}}+K_{\dot{u}}\right)+k\right]\\
 & =\sum_{\tau=k-\matr{T}_{\dot{u}}\left[\dot{m},\dot{l}\right]}^{K_{u}}\hat{h}_{u,\tau,\left(\dot{l}-1\right)\left(N_{\dot{u}}+K_{\dot{u}}\right)+k}\dot{x}_{k-\tau}.\label{eq:diffreceivedsamp}
\end{align}

The \ac{fft} outputs also only need to be updated for a few samples
and taking the \ac{fft} of the whole \ac{ofdm} symbol is not necessary.
Using the previously calculated received symbol estimates, if there
is an update to symbol estimate in the $m$th subcarrier of the $l$th
\ac{ofdm} symbol of $u$th user, the new symbol can be estimated
by adding the contribution from the updated samples and removing the
contribution from the original samples as
\begin{equation}
\widetilde{\matr{Y}_{u}}\left[m,l\right]=\tilde{\matr{Y}}_{u}\left[m,l\right]+\sum_{k=K_{u}+1}^{K_{u}+\matr{T}_{\dot{u}}\left[\dot{m},\dot{l}\right]}\frac{\exp\left(\jmath\tfrac{2\pi M_{u,m}\left(k-K_{u}-1\right)}{N_{u}}\right)}{\sqrt{N_{u}}}\dot{y}_{u,k}.
\end{equation}
The difference in the updated symbol estimate in $u$th user's $l$th
\ac{ofdm} symbol's $m$th subcarrier due to the proposed window is
denoted by
\begin{align}
\dot{Y}_{u,l,m} & =\widetilde{\matr{Y}_{u}}\left[m,l\right]-\tilde{\matr{Y}}_{u}\left[m,l\right]\\
 & =\sum_{k=K_{u}+1}^{K_{u}+\matr{T}_{\dot{u}}\left[\dot{m},\dot{l}\right]}\frac{\exp\left(\jmath\tfrac{2\pi M_{u,m}\left(k-K_{u}-1\right)}{N_{u}}\right)}{\sqrt{N_{u}}}\dot{y}_{u,k}.\label{eq:diffsymbol}
\end{align}

Accordingly, new channel capacity becomes\begin{dgroup}
\begin{dmath}
\widetilde{\matrgr{\eta}_{u}}\left[m,l\right]=\log_{2}\left(1+\frac{\left|\tilde{\matr{H}}_{u}\left[m,l\right]\right|^{2}}{1+\left|\widetilde{\matr{Y}_{u}}\left[m,l\right]-\tilde{\matr{H}}_{u}\left[m,l\right]\mathbf{D}_{u}\left[m,l\right]\right|^{2}}\right)
\end{dmath}
\begin{dmath}
=\log_{2}\left(1+\frac{\left|\tilde{\matr{H}}_{u}\left[m,l\right]\right|^{2}}{1+\left|\dot{Y}_{u,l,m}+\tilde{\matr{Y}}_{u}\left[m,l\right]-\tilde{\matr{H}}_{u}\left[m,l\right]\mathbf{D}_{u}\left[m,l\right]\right|^{2}}\right).\label{eq:diffcap}
\end{dmath}
\end{dgroup}Note that the term $\tilde{\matr{Y}}_{u}\left[m,l\right]-\tilde{\matr{H}}_{u}\left[m,l\right]\mathbf{D}_{u}\left[m,l\right]$
was previously calculated in \prettyref{eq:uncappedcap} and the introduced
difference term can simply be added to the previously calculated sum.
Notice that the capacities of only the \ac{re} of which transmit
pulses overlap with that of the \ac{re} under investigation are changed,
and only these need to be compared. Accordingly, assuming that the
\acp{mcs} are not decided yet, it is to the network's advantage to
transmitter window the \ac{re} under investigation with the according
window duration if the following is positive:
\begin{equation}
\eta_{\Delta}=\prod_{u=1}^{U}\left(\sum_{m=1}^{M_{u}}\widetilde{\matrgr{\eta}}_{u}\left[m,\left\lceil \frac{\dot{l}N_{\dot{u}}}{N_{u}}\right\rceil \right]\right)-\prod_{u=1}^{U}\left(\sum_{m=1}^{M_{u}}\breve{\matrgr{\eta}}_{u}\left[m,\left\lceil \frac{\dot{l}N_{\dot{u}}}{N_{u}}\right\rceil \right]\right).\label{eq:capdiff}
\end{equation}
Or if the \acp{mcs} are already decided, $\eta_{\Delta}$ becomes
\begin{dmath}
\eta_{\Delta}=\prod_{u=1}^{U}\left(\sum_{m=1}^{M_{u}}\min\left(\matr{b}_{u}\left[m,\left\lceil \frac{\dot{l}N_{\dot{u}}}{N_{u}}\right\rceil \right],\widetilde{\matrgr{\eta}}_{u}\left[m,\left\lceil \frac{\dot{l}N_{\dot{u}}}{N_{u}}\right\rceil \right]\right)\right)-\prod_{u=1}^{U}\left(\sum_{m=1}^{M_{u}}\min\left(\matr{b}_{u}\left[m,\left\lceil \frac{\dot{l}N_{\dot{u}}}{N_{u}}\right\rceil \right],\breve{\matrgr{\eta}}_{u}\left[m,\left\lceil \frac{\dot{l}N_{\dot{u}}}{N_{u}}\right\rceil \right]\right)\right).\label{eq:boundedcapdiff}
\end{dmath}

Consequently, Algorithm 1 is proposed to iteratively calculate the
optimum transmitter windowing duration at the \ac{gnb}. 
\begin{figure}
\begin{algorithmic}[1]
\Algphase{Algorithm 1}{Estimate $\matr{T}_{u},\,\forall u\in\mathbb{N}_{\leq U}^{*}$ \& Calculate $\matr{x}$}
\State $\matr{T}_u \gets 0,\ \forall u \in \mathbb{N}^{+}_{\leq U} $
\State $ \breve{\matr{x}} \gets $ \prettyref{eq:basetxsymgen}
\ForAll{$u \in \mathbb{N}_{\leq U}^*,\, \tau\in \mathbb{N}_{\leq K_u},\, t \in \mathbb{N}_{\leq \left(N_u+K_u\right)L_u}^{*}$}
\State Predict \ac{dl} \acp{cir} $\tilde{\matr{h}}_{u,\tau,t}$ and \acp{cfr} $\tilde{\matr{H}}_{u}$
\EndFor
\ForAll{$u \in \mathbb{N}_{\leq U}^*,\, m \in \mathbb{N}_{\leq M_u}^{*},\, l \in \mathbb{N}_{\leq L_u}^{*}$}
\State Calculate \cref{eq:estimatedreceivedsamples,eq:estimatedreceivedsymbols,eq:uncappedcap}
\State $ \matrgr{\lambda}_{u}\left[m,l\right] \gets $ \prettyref{eq:uncappedcap}
\If{\acp{mcs} fixed}
\State $ \matrgr{\lambda}_{u}\left[m,l\right] \gets \matrgr{\lambda}_{u}\left[m,l\right]-\matr{b}_u \left[m,l\right] $
\EndIf
\EndFor
\For{$ \dot{m},\dot{l},\dot{u} \gets \argmax\limits_{m,l,u} \matrgr{\lambda}_{u}\left[m,l\right] , \argmin\limits_{m,l,u} \matrgr{\lambda}_{u}\left[m,l\right] $}
\For{$\matr{T}_{\dot{u}}\left[\dot{m},\dot{l}\right]\gets 1,K_u$}
\State $\eta_\Delta \gets $ \prettyref{eq:capdiff} or \prettyref{eq:boundedcapdiff} \label{lst:line:getcapdiff}
\If{$\eta_\Delta>0$}
\ForAll{$u \in \mathbb{N}_{\leq U}^{*}, m \in \mathbb{N}_{\leq M_u}^{*}$}
\State $ \breve{\matrgr{\eta}}_{u}\left[m,\left\lceil \frac{\dot{l}N_{\dot{u}}}{N_{u}}\right\rceil \right] \gets \widetilde{\matrgr{\eta}}_{u}\left[m,\left\lceil \frac{\dot{l}N_{\dot{u}}}{N_{u}}\right\rceil \right] $
\EndFor
\Else
\State $ \matr{T}_{\dot{u}}\left[\dot{m},\dot{l}\right] \gets \matr{T}_{\dot{u}}\left[\dot{m},\dot{l}\right]-1 $
\If{$\matr{T}_{\dot{u}}\left[\dot{m},\dot{l}\right]>0$} 
\ForAll{$k \in \mathbb{N}^{*}_{\leq \matr{T}_{\dot{u}}\left[\dot{m},\dot{l}\right]}$}
\State $ \breve{\matr{x}}\left[\left(\dot{l}-1\right)\left(N_{\dot{u}}+K_{\dot{u}}\right)+k\right] \gets $ \prettyref{eq:txwinconvertsamp}
\EndFor
\EndIf
\State \textbf{break}
\EndIf
\EndFor
\EndFor
\end{algorithmic}
\end{figure}
The variable introduced in Algorithm 1, $\matrgr{\lambda}_{u}\in\mathbb{R}^{M_{u}\times L_{u}}$,
corresponds to the excess \ac{snr} of the \ac{re} if \acp{mcs}
are determined, or to the \ac{snr} of the \ac{re} if not. On par
with the motivation behind Algorithm 1, the \acp{re} that have higher
excess \ac{snr} are more likely to have longer windowing durations
resulting in more significant overall interference reduction before
those with lesser impact are pursued. Since there is no additional
extension to \ac{cp}, which is currently designed only to support
the multipath channel, all \acp{re} are assumed to have a zero transmitter
windowing duration initially. The duration is incremented instead
of a binary search as the expected window durations are short and
calculation of shorter durations are computationally less exhaustive
as will be described in \prettyref{subsec:Computational-Complexity-txwin}.
The algorithm is provided in a recursive manner for brevity, but the
equations invoked by \prettyref{eq:capdiff}/\prettyref{eq:boundedcapdiff}.
It should also be noted that Algorithm 1 runs only at the \ac{gnb}
which virtually has no computational complexity and power limitations
while the \acp{ue} are unaware of the process and are not passed
any information. This makes Algorithm 1 forward and backward compatible
with all communication standards.

\subsubsection{Computational Complexity\label{subsec:Computational-Complexity-txwin}}

Channel prediction, and mean \ac{snr} and capacity estimation for
each user is assumed to be performed for link adaptation purposes\citep{oien2004impactof}
regardless of Algorithm 1 and is not considered in the computational
complexity of proposed algorithm. There are many computational complexity
reducing implementation tricks used in \prettyref{subsec:Transmitter-Algorithm}.
The computational complexity of the algorithm is derived by counting
the number of operations performed for each step and how many times
those steps were invoked. \prettyref{tab:Computational-Complexity-txwin}
shows the number of real additions and multiplications required to
test whether windowing an \ac{re} at the transmitter with a duration
of $T$ is beneficial, i.e. executing line~\ref{lst:line:getcapdiff}
of Algorithm 1, how many times each equation is invoked, and the total
number of necessary operations. It is shown that each test requires
\prettyref{eq:capdiffcomplexitysum} real additions and \prettyref{eq:capdiffcomplexitymult}
real multiplications. Accordingly determining the optimum transmitter
window durations for all \acp{re} in the transport block, and windowing
the sample sequence accordingly results in $\sum_{u=1}^{U}\sum_{l=1}^{L_{u}}\sum_{m=1}^{M_{u}}\left(2\matr{T}_{u}\left[m,l\right]+\sum_{T=1}^{\min\left(\matr{T}_{u}\left[m,l\right]+1,K_{u}\right)}\prettyref{eq:capdiffcomplexitysum}\right)$
real additions and $\sum_{u=1}^{U}\sum_{l=1}^{L_{u}}\sum_{m=1}^{M_{u}}\sum_{T=1}^{\min\left(\matr{T}_{u}\left[m,l\right]+1,K_{u}\right)}\prettyref{eq:capdiffcomplexitymult}$
real multiplications. Other than this, Algorithm 1 also needs the
calculate the fair proportional network capacity for the nonwindowed
case, requiring $\sum_{u=1}^{U}\left(M_{u}-1\right)L_{u}$ real additions,
and $\max L_{u}$ real multiplications if there are only 2 different
subcarrier spacings or $\frac{3}{2}\max L_{u}$ real multiplications
if all three subcarrier spacing possibilities for the band is used.
Statistics regarding the distribution of $T$ and according number
of calculations for the evaluated scenarios are provided in \prettyref{sec:Numerical-Verification}.
Regarding the timewise complexity, it should be noted that the calculation
can be done in parallel for the $\min L_{u}$ independent symbol groups,
and therefore the worst-case time complexity of the described computationally
efficient implementation is $\mathcal{O}\left(\sum_{u}M_{u}K_{u}^{2}\right)$,
whereas a more operation count- and memory-wise exhaustive implementation
can complete in $\mathcal{O}\left(\max\left(K_{u}\right)+\max\left(M_{u}\right)\right)$,
which may be feasible at the \ac{gnb}. Further operational and timewise
computational complexity reduction can be obtained if the Algorithm
is only run for a subset of subcarriers such as \citep{Sahin2011}.
\begin{table}
\caption{Computational Complexity of Each Call to \prettyref{eq:capdiff}/\prettyref{eq:boundedcapdiff}\label{tab:Computational-Complexity-txwin}}

\centering%
\begin{tabular}{>{\centering}m{0.05\columnwidth}|>{\centering}m{0.31\columnwidth}|>{\centering}m{0.33\columnwidth}|c}
Eq. & \#Add & \#Mult. & \#Inv.\tabularnewline
\hline 
\hline 
\prettyref{eq:txwinsampdiff} & $6T$ & $10T$ & 1\tabularnewline
\hline 
\prettyref{eq:diffreceivedsamp} & $2T^{2}$ & $2T^{2}+2T$ & $U$\tabularnewline
\hline 
\prettyref{eq:diffsymbol} & $4T-2$ & $4T$ & $\sum_{u=1}^{U}M_{u}$\tabularnewline
\hline 
\prettyref{eq:diffcap} & 3 & 3 & $\sum_{u=1}^{U}M_{u}$\tabularnewline
\hline 
\prettyref{eq:capdiff} /\prettyref{eq:boundedcapdiff} & $1+\sum_{u=1}^{U}\left(M_{u}-1\right)$ & $U-1$ & 1\tabularnewline
\hline 
\hline 
Total & \begin{dmath}
2UT^{2}+T\left(4\sum_{u=1}^{U}M_{u}+6\right)+2\sum_{u=1}^{U}M_{u}+1-U\label{eq:capdiffcomplexitysum}
\end{dmath} & \begin{dmath}
2UT^{2}+T\left(2U+4\sum_{u=1}^{U}M_{u}+10\right)+U+3\sum_{u=1}^{U}M_{u}-1\label{eq:capdiffcomplexitymult}
\end{dmath} & \tabularnewline
\end{tabular}
\end{table}

\subsection{Estimation of Optimum Receiver Window Durations\label{subsec:Optimum-Receiver-Window}}

A theoretical approach requiring knowledge regarding channel conditions
of at least the \acp{ue} utilizing adjacent bands was proposed in
\citep{pekoz2017adaptive}. Although the approach in \citep{pekoz2017adaptive}
is theoretically optimal, it is not feasible for use especially in
the \ac{dl} due to the extent of required data (at least \acp{pdp},
or better yet, \acp{cir} between the transmitters of signals occupying
adjacent bands and the receiver) at the \acp{ue}. In this work, we
propose calculating receiver window duration solely using the statistics
of the received signal. Sole dependence on statistics allows each
\ac{ue} to perform their own estimation in a decentralized manner
without the need for network-wide channel and data knowledge required
in \citep{pekoz2017adaptive}. Since calculations are done only by
the intended receiver and receiver windowing only affects the \ac{sinr}
of the \ac{re} that the operation is applied to, there is no need
to convey any information to and from other nodes and maximization
of fair-proportional network capacity is achieved by independently
maximizing the capacity of each \ac{re}. This makes the proposed
algorithm backward and forward compatible with any communication standard
and protocol. Furthermore, computationally efficient receiver windowing
of \ac{ofdm} symbols for multiple receiver window durations are discussed
and the computational complexity of the proposed technique is derived.

\subsubsection{Computationally Efficient Conversion of Conventionally Received \ac{cp}-\ac{ofdm}
Symbols to Per-\ac{re} Receiver Windowed \ac{ofdm} Symbols}

\begin{figure}
\includegraphics{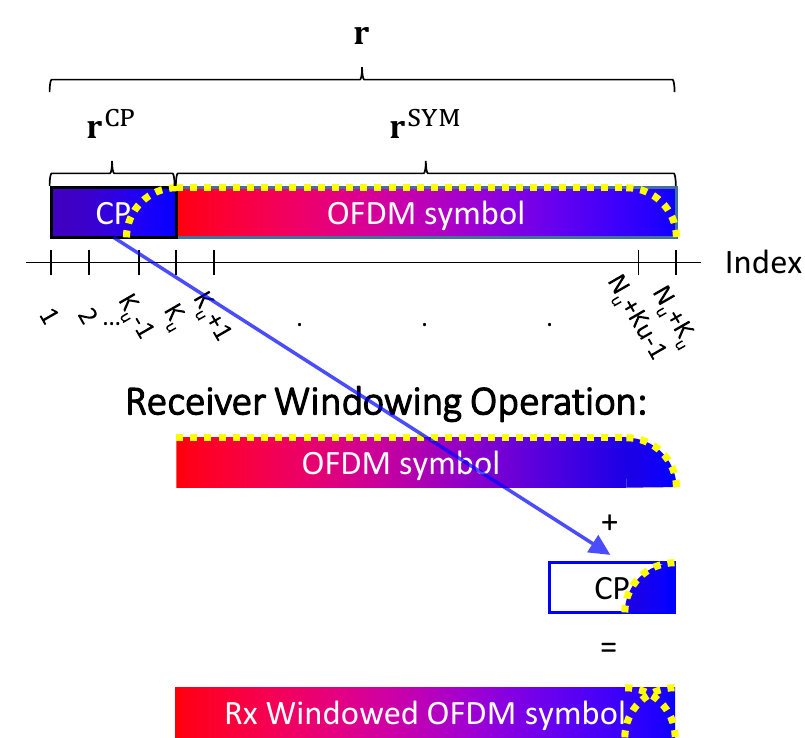}

\caption{Indexing of $\matr{r}$ and identification of its parts $\matr{r}^{\text{CP}}$
and $\matr{r}^{\text{SYM}}$ within a demonstration of how receiver
windowing operation is performed.\label{fig:Fig3}}
\end{figure}

Assume $u$th \ac{ue} uses the receiver windowing pulse shape $\matr{r}_{m,l,u}\in\mathbb{R}^{\left(N_{u}+K_{u}\right)\times1}$
of which indexing is shown in \prettyref{fig:Fig3} calculated according
to \citep{guvenkaya2015awindowing} to reject the energy outside the
\ac{ue}'s band with a receiver windowing duration of $\matr{R}_{u}\left[m,l\right]$
to receive the $m$th subcarrier of $l$th \ac{ofdm} symbol. As also
discussed in \citep{guvenkaya2015awindowing}, a visual investigation
of \prettyref{fig:Fig3} reveals that the analyzed receiver windowed
single carrier symbols differ from that of the \ac{fft} output by
the last $\matr{R}_{u}\left[m,l\right]$ samples. The contribution
from the $s\in\mathbb{N}_{K_{u}<s\leq K_{u}+N_{u}}^{*}$th sample
to the \ac{fft} output, if windowing is not performed, is, $\matr{y}_{l,u}\left[s\right]\exp\left(\jmath2\pi M_{u,m}\left(s-K_{u}-1\right)/N_{u}\right)/\sqrt{N_{u}}$.
If windowing is applied, for $s\in\mathbb{N}_{N_{u}<s\leq K_{u}+N_{u}}^{*}$,
the contribution instead becomes
\begin{dmath}[label={rxwinsampcontrib}]
\left(\matr{y}_{l,u}\left[s\right]\matr{r}_{m,l,u}\left[s\right]+\matr{y}_{l,u}\left[s-N_{u}\right]\matr{r}_{m,l,u}\left[s-N_{u}\right]\right)\tfrac{\exp\left(\jmath\tfrac{2\pi M_{u,m}\left(s-K_{u}-1\right)}{N_{u}}\right)}{\sqrt{N_{u}}}.
\end{dmath}Accordingly, by removing the non-windowed contribution from all windowed
samples and adding their respective windowed contribution to the \ac{fft}
output, the \ac{sc} symbol that is receiver windowed with window
duration $0<r\leq K_{u}$ can be written as \begin{dmath}[label={rxwinsymlong}]
\matr{Y}_{u}\left[m,l,r\right]=\matr{Y}_{u}\left[m,l,0\right]+\sum_{s=N_{u}+K_{u}-r+1}^{N_{u}+K_{u}}\left(\matr{y}_{l,u}\left[s\right]\left(\matr{r}_{m,l,u}\left[s\right]-1\right)+\matr{y}_{l,u}\left[s-N_{u}\right]\matr{r}_{m,l,u}\left[s-N_{u}\right]\right)\tfrac{\exp\left(\jmath\tfrac{2\pi M_{u,m}\left(s-K_{u}-1\right)}{N_{u}}\right)}{\sqrt{N_{u}}}.
\end{dmath}Plugging $\matr{r}_{m,l,u}\left[s\right]=1-\matr{r}_{m,l,u}\left[s-N_{u}\right]$
for the windowed region per \citep{guvenkaya2015awindowing,bala_shaping_2013},
\prettyref{eq:rxwinsymlong} can be simplified to\begin{dmath}[label={rxwinsym}]
\matr{Y}_{u}\left[m,l,r\right]=\matr{Y}_{u}\left[m,l,0\right]+\sum_{s=N_{u}+K_{u}-r+1}^{N_{u}+K_{u}}\left(\matr{y}_{l,u}\left[s-N_{u}\right]-\matr{y}_{l,u}\left[s\right]\right)\matr{r}_{m,l,u}\left[s-N_{u}\right]\tfrac{\exp\left(\jmath\tfrac{2\pi M_{u,m}\left(s-K_{u}-1\right)}{N_{u}}\right)}{\sqrt{N_{u}}},
\end{dmath}which allows computing the receiver windowed symbols with reduced
computational complexity.

\subsubsection{Optimum Receiver Windowing Duration Estimation Algorithm}

The optimum receiver windowing duration similarly maximizes \prettyref{eq:uncappedcap}.
However, unlike the \ac{gnb} that has predicted the \ac{cfr} coefficients
and already knows the payload, the \acp{ue} know neither. However,
there are other higher order statistics that can be exploited by the
\acp{ue}. Similar to \prettyref{eq:estimatedreceivedsymbols}, one
can write\begin{dmath}
\matr{Y}_{u}\left[m,l,r\right]=\matr{H}_{u}\left[m,l\right]\left(\matr{D}_{u}\left[m,l\right]+\matr{P}_{u}\left[m,l\right]\right)+\tilde{\matr{N}}_{u}\left[m,l,r\right]+\tilde{n}_{u}\left[m,l,r\right],
\end{dmath}where $\matr{H}_{u}\left[m,l\right]$ is the actual \ac{cfr} coefficient
affecting the $m$th subcarrier of $l$th \ac{ofdm} symbol of $u$th
user, $\tilde{\matr{N}}_{u}\left[m,l,r\right]$ is the combined \ac{aci},
\ac{ici} and \ac{isi}\footnote{Although this element consists of the sum of each of these components
scaled with different coefficients, all varying with used window,
only this combined element will be referred to for the sake of brevity
as future analysis only involves the sum.} affecting the aforementioned \ac{re} if receiver window duration
$r$ is used, and $\tilde{n}_{u}\left[m,l,r\right]$ is the noise
value affecting the aforementioned \ac{re}. Let the 2-tuple elements
of the set $\mathbb{P}_{u}\left[\dot{m},\dot{l}\right]$ refer to
the subcarrier and \ac{ofdm} symbol indices of $P$ \acp{re} that
are statistically expected to experience the channels most correlated
with $\matr{H}_{u}\left[\dot{m},\dot{l}\right]$\citep{bello1963characterization}.
To keep equations concise, we will only use $\matr{D}_{u}\left[m,l\right]$
to refer to $\matr{D}_{u}\left[m,l\right]+\matr{P}_{u}\left[m,l\right]$
from this point onward. Even though no element other than $\matr{P}_{u}\left[m,l\right]$
in the equation is known, the \ac{ue} can still obtain\begin{dgroup}
\begin{dmath}
\breve{\matr{Y}}_{u}\left[\dot{m},\dot{l},r\right]\coloneqq\var{\left\{ \matr{Y}_{u}\left[m,l,r\right],\left(m,l\right)\in\mathbb{P}_{u}\left[\dot{m},\dot{l}\right]\right\} }
\end{dmath}\begin{dmath}
=\var{\left\{ \matr{H}_{u}\left[m,l\right]\matr{D}_{u}\left[m,l\right]+\tilde{\matr{N}}_{u}\left[m,l,r\right]+\tilde{n}_{u}\left[m,l,r\right]\right\} }
\end{dmath}\begin{dmath}
=\var{\left\{ \matr{H}_{u}\left[m,l\right]\matr{D}_{u}\left[m,l\right]\right\} }+\var{\left\{ \tilde{n}_{u}\left[m,l,r\right]\right\} }+\var{\left\{ \tilde{\matr{N}}_{u}\left[m,l,r\right]\right\} },
\end{dmath}
\end{dgroup}where the set definitions $\left(m,l\right)\in\mathbb{P}_{u}\left[\dot{m},\dot{l}\right]$
were removed after the first line to keep equations concise, but are
always implied throughout the rest of this section for all mean and
variance operations, and an equal-weight variance is assumed, or in
probability terms, all elements are assigned the same $1/P$ probability.
Weighting elements with the correlation between $\matr{H}_{u}\left[m,l\right]$
and $\matr{H}_{u}\left[\dot{m},\dot{l}\right]$\citep{clarke1968astatistical}
is optimum\citep{kahn1954ratiosquarer}, however, the equiweight implementation
drastically reduces the computational complexity as will be shown
below, without an observable performance loss. Note that since $\tilde{n}_{u}\left[m,l,r\right]\sim\mathcal{CN}\left(0,1\right)\forall u,m,l,r$,
although the noise value itself changes with windowing, the noise
variance remains unity. Furhermore, as \ac{ici} and \ac{isi} are
separated, the variance in the actual channel coefficients can be
assumed to remain constant regardless of window duration as well.
Thus, the \ac{cfr} coefficient, transmitted data and noise variance
remain constant regardless of applied window, but the combined interference
and its variance varies with the windowing operation. Although it
is impossible to distinguish between these components by looking at
the effects of windowing on a single received symbol, the spectrotemporal
correlation of channel and interference can be exploited to identify
the amount of combined interference in a group of \acp{re}. That
is, although $\var{\left\{ \tilde{\matr{N}}_{u}\left[m,l,r\right],\left(m,l\right)\in\mathbb{P}_{u}\left[\dot{m},\dot{l}\right]\right\} }$
can not be found explicitly, one can conclude that
\begin{equation}
\argmin_{r}\breve{\matr{Y}}_{u}\left[\dot{m},\dot{l},r\right]\triangleq\argmin_{r}\var{\left\{ \tilde{\matr{N}}_{u}\left[m,l,r\right]\right\} }.\label{eq:rxwinoptcond}
\end{equation}

The optimum receiver windowing duration calculation algorithm utilizes
\prettyref{eq:rxwinoptcond} to minimize the combined interference
energy and maximize capacity. With similar reasoning to Algorithm
1, Algorithm 2 also starts with the assumption of zero initial window
duration, and checks to see whether longer window durations are beneficial
for each \ac{re}. Let us now investigate a possible reduced complexity
implementation of this idea, particularly utilizing the relation between
$\matr{Y}_{u}\left[m,l,0\right]$ and $\matr{Y}_{u}\left[m,l,r\right]$
as shown in \prettyref{eq:rxwinsymlong}. Let us first define
\begin{figure}
\begin{algorithmic}[1]
\Algphase{Algorithm 2}{Estimate $\matr{R}_{u} $ \& $  \hat{\matr{Y}}_u$}
\State $\matr{R}_{u}\gets 0$
\ForAll{$ m \in M_{u}, l \in L_{u} $}
\State $\breve{\matr{Y}}_{u}\left[m,l,0\right] \gets \prettyref{eq:rxsymvariancedef}$
\For{$r\gets 1,K_{u}$}
\State $\breve{\matr{Y}}_{u}\left[m,l,r\right] \gets \prettyref{eq:rxwinvariance}$
\If{$\breve{\matr{Y}}_{u}\left[m,l,r\right]>\breve{\matr{Y}}_{u}\left[m,l,r-1 \right]$}
\State $\matr{R}_u\left[m,l\right]\gets r-1$
\State \textbf{break}
\EndIf
\EndFor
\State $ \hat{\matr{Y}}_u\left[ m,l \right] \gets \prettyref{eq:rxwinsym} $
\EndFor
\end{algorithmic}
\end{figure}
\begin{dgroup}
\begin{dmath}
\ddot{y}_{u}\left[m,l,r\right]=\matr{Y}_{u}\left[m,l,r\right]-\matr{Y}_{u}\left[m,l,0\right]
\end{dmath}\begin{dmath}
=\sum_{s=N_{u}+K_{u}-r+1}^{N_{u}+K_{u}}\left(\matr{y}_{l,u}\left[s-N_{u}\right]-\matr{y}_{l,u}\left[s\right]\right)\matr{r}_{m,l,u}\left[s-N_{u}\right]\tfrac{\exp\left(\jmath\tfrac{2\pi M_{u,m}\left(s-K_{u}-1\right)}{N_{u}}\right)}{\sqrt{N_{u}}}\label{eq:rxwinsymdiff}
\end{dmath}
\end{dgroup}to keep following equations concise. Then\begin{dgroup}
\begin{dmath}
\breve{\matr{Y}}_{u}\left[\dot{m},\dot{l},r\right]=\sum_{\left(m,l\right)\in\mathbb{P}_{u}\left[\dot{m},\dot{l}\right]}\frac{\left|\matr{Y}_{u}\left[m,l,r\right]-\sum_{\left(\ddot{m},\ddot{l}\right)\in\mathbb{P}_{u}\left[\dot{m},\dot{l}\right]}\matr{Y}_{u}\left[\dot{m},\dot{l},r\right]/P\right|^{2}}{P}
\end{dmath}\begin{dmath}
=\frac{1}{P^{3}}\sum_{\left(m,l\right)\in\mathbb{P}_{u}\left[\dot{m},\dot{l}\right]}\left|\matr{Y}_{u}\left[m,l,r\right]\left(P-1\right)-\sum_{\left(\ddot{m},\ddot{l}\right)\in\mathbb{P}_{u}\left[\dot{m},\dot{l}\right]\setminus\left(m,l\right)}\matr{Y}_{u}\left[\ddot{m},\ddot{l},r\right]\right|^{2}\label{eq:rxsymvariancedef}
\end{dmath}\begin{dmath}
=\frac{1}{P^{3}}\sum_{\left(m,l\right)\in\mathbb{P}_{u}\left[\dot{m},\dot{l}\right]}\left|\matr{Y}_{u}\left[m,l,0\right]\left(P-1\right)-\sum_{\left(\ddot{m},\ddot{l}\right)\in\mathbb{P}_{u}\left[\dot{m},\dot{l}\right]\setminus\left(m,l\right)}\matr{Y}_{u}\left[\ddot{m},\ddot{l},0\right]+\ddot{y}_{u}\left[m,l,r\right]\left(P-1\right)-\sum_{\left(\ddot{m},\ddot{l}\right)\in\mathbb{P}_{u}\left[\dot{m},\dot{l}\right]\setminus\left(m,l\right)}\ddot{y}_{u}\left[\ddot{m},\ddot{l},r\right]\right|^{2}\label{eq:rxwinvariance}
\end{dmath}
\end{dgroup}demonstrates that once 
\begin{equation}
\matr{Y}_{u}\left[m,l,0\right]\left(P-1\right)-\sum_{\left(\ddot{m},\ddot{l}\right)\in\mathbb{P}_{u}\left[\dot{m},\dot{l}\right]\setminus\left(m,l\right)}\matr{Y}_{u}\left[\ddot{m},\ddot{l},0\right]\label{eq:meandiff}
\end{equation}
 is calculated, the variance of the windowed cases can be calculated
by adding the window differences and summing the squared magnitudes.
The advantage of the equiweight assumption becomes clear at this point,
a simple investigation reveals that once $\breve{\matr{Y}}_{u}\left[\dot{m},\dot{l},r\right]$
is calculated for an \ac{re}, the same calculation for neighboring
\acp{re} only require adding and removing contributions from few
\acp{re}. More information on computational complexity is provided
in \prettyref{subsec:Computational-Complexity-rxwin}.

\subsubsection{Computational Complexity\label{subsec:Computational-Complexity-rxwin}}

Calculating \prettyref{eq:meandiff} for $\left(m,l\right)\gets\left(m_{1},l_{1}\right)\in\mathbb{P}_{u}\left[\dot{m},\dot{l}\right]$
for a single \ac{re} requires $2P$ real additions and $2$ real
multiplications. The result of the same equation for another \ac{re}
with indices $\left(m,l\right)\gets\left(m_{2},l_{2}\right)\in\mathbb{P}_{u}\left[\dot{m},\dot{l}\right]$
can be obtained by adding $P\left(\matr{Y}_{u}\left[m_{2},l_{2},0\right]-\matr{Y}_{u}\left[m_{1},l_{1},0\right]\right)$
to the previously calculated value, resulting in $4$ real additions
and $2$ real multiplications. Thus, calculating \prettyref{eq:meandiff}
$\forall\left(m,l\right)\in\mathbb{P}_{u}\left[\dot{m},\dot{l}\right]$
requires a total of $6P-4$ real additions and $2P$ real multiplications.

Trials show that the subsets $\mathbb{P}_{u}\left[\dot{m},\dot{l}\right]$
differ at most by $\log\left(c+P\right)$ individual \acp{re} for
neighbor \acp{re} under vehicular channels\citep{3gpp.38.901} for
statistically meaningful $P$ values, where $c$ is a small positive
constant. While the mean subset difference is well below that for
the possible \ac{tti} durations and bandwidth part configurations
in \ac{nr}, $\log\left(P\right)$ will be assumed for all \acp{re}
as the mean asymptotically reaches this number with increasing number
of allocated slots and \acp{rb}, and to mitigate $c$. Thus, after
\prettyref{eq:meandiff} is calculated for an \ac{re} for $\mathbb{P}_{u}\left[\dot{m}_{1},\dot{l}_{1}\right]$,
the results can be generalized for the same \ac{re} for another $\mathbb{P}_{u}\left[\dot{m}_{2},\dot{l}_{2}\right],\left(\dot{m}_{2},\dot{l}_{2}\right)\gets\exists\left\{ \left(\dot{m}_{1}\pm1,\dot{l}_{1}\right),\left(\dot{m}_{1},\dot{l}_{1}\pm1\right)\right\} $
by adding $P\left(\sum_{\left(\ddot{m},\ddot{l}\right)\in\mathbb{P}_{u}\left[\dot{m}_{2},\dot{l}_{2}\right]}\matr{Y}_{u}\left[\ddot{m},\ddot{l},0\right]-\sum_{\left(\ddot{m},\ddot{l}\right)\in\mathbb{P}_{u}\left[\dot{m}_{1},\dot{l}_{1}\right]}\matr{Y}_{u}\left[\ddot{m},\ddot{l},0\right]\right)$
to the previous findings, which requires $4\log\left(P\right)$ real
additions and $2$ real multiplications. The findings can then similarly
propagate to other \acp{re} $\in\mathbb{P}_{u}\left[\dot{m}_{2},\dot{l}_{2}\right]$
by performing $4$ real additions and $2$ real multiplications each
as described above. Therefore, number of operations required to obtain
\prettyref{eq:meandiff} $\forall\left(m,l\right),\forall\left(\dot{m},\dot{l}\right)$
is upper bounded by $4\left(M_{u}L_{u}\left(P+\log\left(P\right)-1\right)-\log\left(P\right)\right)+2P$
real additions and $2PM_{u}L_{u}$ real multiplications.

A direct investigation reveals that each \prettyref{eq:rxwinsymdiff}
calculation requires $6r-2$ real additions and $6r$ real multiplications
to obtain the symbol windowed with window duration $r$. Once the
relevant \prettyref{eq:rxwinsymdiff} values are calculated, the number
of equations required to calculate the difference $\ddot{y}_{u}\left[m,l,r\right]\left(P-1\right)-\sum_{\left(\ddot{m},\ddot{l}\right)\in\mathbb{P}_{u}\left[\dot{m},\dot{l}\right]\setminus\left(m,l\right)}\ddot{y}_{u}\left[\ddot{m},\ddot{l},r\right]$
in \prettyref{eq:rxwinvariance} is the same as the number of operations
required to obtain \prettyref{eq:meandiff}. It should be noted that
these values are only required for $\left(m,l\right)\in\mathbb{P}_{u}\left[\dot{m},\dot{l}\right]$
if $\breve{\matr{Y}}_{u}\left[\dot{m},\dot{l},r\right]$ is being
calculated, which is not always needed.

After both differences in \prettyref{eq:rxwinvariance} is obtained,
the sum of the squared magnitudes of the sum of differences can be
calculated to finalize \prettyref{eq:rxwinvariance} calculation.
This requires $3P-1$ real additions and $2P$ real multiplications.
If $\matr{R}_{u}\left[m,l\right]=R$, \prettyref{eq:rxwinvariance}
must be calculated $\forall r\in\mathbb{N}_{\leq\min\left(R+1,K_{u}\right)}^{*}$.
Once $\matr{R}_{u}$ is found, \prettyref{eq:rxwinsym} is performed
to obtain windowed symbols to continue reception, which requires only
$2\#\left\{ \matr{R}_{u}\neq0\right\} $ real additions and no multiplications.
Some statistics for $\matr{R}_{u}$ and number of operations performed
for vehicular channel conditions are provided in \prettyref{sec:Numerical-Verification}.
It should also be noted that the worst case time complexity of the
described efficient implementation is on the order of $\mathcal{O}\left(K_{u}^{2}PM_{u}L_{u}\right)$,
while a more straightforward operation count- and memory-wise exhaustive
implementation can run within $\mathcal{O}\left(P+K_{u}\right)$.

\subsection{Further Notes on Computational Complexity}

The algorithms presented in \prettyref{sec:Proposed-Method} are computationally
tailored around the basic assumption that both transmitter and receiver
window durations are expected to be short as the utilized extension
was solely intended for the channel. While \prettyref{sec:Numerical-Verification}
shows that this assumption holds, there are also other characteristics
that can be exploited, such as the spectrotemporal correlation of
window durations, and a non-obvious but comprehensible peak in the
statistical receiver window duration probability distribution, all
of which are presented and discussed in \prettyref{sec:Numerical-Verification}.
This section was aimed to describe the basic ideas and only simple,
universal algorithmic implementation specific details in the most
comprehensible manner. Further possible reductions in computational
complexity are mentioned along with numerical findings in \prettyref{sec:Numerical-Verification}. 

\section{Numerical Verification\label{sec:Numerical-Verification}}

Although the proposed method is formulated for networks with any number
of \acp{ue}, in this work, a simple network limited to a \ac{bs}
and two \acp{ue} equally sharing a $\SI{7.68}{\mega\hertz}$ system
bandwidth is considered for the sake of simplicity, as done in other
similar works such as \citep{andrews2001providing}. This also allows
clearer presentation of the results. This network is realized numerous
times with independent and random user data and instantaneous channels,
and all presented results are the arithmetic means of all realizations
unless otherwise specified. The parameters provided in \citep{3gpp.38.802}
for link level waveform evaluation under $\SI{6}{\giga\hertz}$ were
used when possible. One of the \acp{ue} is a high mobility node experiencing
a channel that has $\SI{30}{\nano\second}$ RMS delay spread and $\SI[per-mode=symbol]{120}{\kilo\meter\per\hour}$
mobility, hereinafter referred to as the ``'f'ast user'', communicating
using 60 subcarriers of an \ac{ofdm} numerology with subcarrier spacing
of $\Delta_{f_{f}}=\SI{60}{\kilo\hertz}$. The second \acp{ue} is
a moderate mobility node experiencing a channel that has $\SI{100}{\nano\second}$
RMS delay spread and $\SI[per-mode=symbol]{30}{\kilo\meter\per\hour}$
mobility, hereinafter referred to as the ``'s'low user'', communicating
using 120 subcarriers of the $\Delta_{f_{s}}=\SI{30}{\kilo\hertz}$
numerology in the adjacent band. The \ac{pdp} of fast user's channel
is \ac{3gpp} \ac{tdl}-A\citep{3gpp.38.901} in $\nicefrac{1}{2}$,
\ac{tdl}-B in $\nicefrac{1}{3}$ and \ac{tdl}-C in $\nicefrac{1}{6}$
of the simulations to demonstrate the operability of the algorithm
under different channel models. Similarly, the \ac{pdp} of slow user's
channel is \ac{3gpp} \ac{tdl}-B in $\nicefrac{1}{2}$, and \ac{tdl}-A
or \ac{tdl}-C each in $\nicefrac{1}{4}$ of the simulations. The
Doppler spectra of both channels are assumed to be classical Jakes
\citep{gans1972apowerspectral} at all times\citep{3gpp.38.901}.
There is a $\SI{240}{\kilo\hertz}$ guard band between users. The
\ac{snr} of each user is sweeped from \SIrange{5}{15}{\deci\bel}
during which the \ac{snr} of the other user is fixed to $\SI{10}{\deci\bel}$. 

Results are obtained for a duration of one \ac{nr} format 4 slot\citep{3gpp.38.213}
in the slow user's reference, where both flexible symbols are utilized
for \ac{ul}. The \ac{ul} transmission interval of a slot followed
by the \ac{dl} transmission interval of the consecutive slot is investigated.
There's a timing offset of 64 samples in the \ac{ul}, whereas the
consequent \ac{dl} period is synchronous. The \ac{ul} \ac{dmrs}
received at the \ac{gnb}, which are \ac{pusch} \ac{dmrs} type B
mapped\citep{3gpp.38.211}, are used to estimate the channel. Only
this time invariant estimate is used in Algorithm 2 for the following
\ac{dl} transmission interval. This presents the worst-case performance
of especially Algorithm 2 under minimum available information. The
rate of performance improvement for increasing number of consecutive
slots with the help of channel prediction\citep{oien2004impactof}
is left for future work. The \ac{dl} \ac{dmrs} configuration is
single port single layer mapped with crucial parameters uniquely defining
the mapping dmrs-AdditionalPosition 3 and dmrs-TypeA-Position pos2\citep{3gpp.38.211}.
No windowing or power control is applied to \ac{ul} signals as well,
reducing the performance of solely the proposed methods making it
the worst case scenario.

Unless otherwise specified, both \acp{ue} utilize a normal \ac{cp}
overhead of $\nicefrac{9}{128}$ with no additional extension for
windowing at all times, thus conserving standard 5G \ac{nr} symbol
structure. For comparison, optimum fixed extension windowing algorithm\citep{guvenkaya2015awindowing}
is also featured utilizing the standard extended \ac{cp} overhead
of 25\% and the additional extension is used for either transmitter
or receiver windowing, as well as \ac{fofdm}\citep{abdoli_filtered_2015,HuaweiFOFDM2016}
and \ac{ncofdm}\citep{beek2009ncontinuous}, the tone offset for
the former, in accord with the resource allocation, being 7.5 and
3.5 tones for the slow and fast user, respectively; and the $N$ parameter
for the latter being $N_{\text{fast}}=1$ and $N_{\text{slow}}=2$
per the original work, and both receivers use the iterative correcting
receiver\citep[Sec. 3]{beek2009ncontinuous} performing 8 iterations.
Link adaptation is omitted in the system, all \acp{rb} are assigned
the same constant \ac{mcs} which consists of QPSK modulation and
$\left(21/32\right)\times\left(7/15\right)$ standard\citep{Bose1960,Hocquenghem1959}
and extended\citep{peterson1965onthe} \ac{bch} \acl{tpc} \citep{pyndiah1998nearoptimum}
for the slow user and $\left(7/16\right)\times\left(7/15\right)$
for the fast user at all times. The \acp{mcs} are chosen such that
both users operate slightly below the target \ac{ber} at the minimum
\ac{snr}, thus the \ac{snr} difference between the users can be
referred to as the excess \ac{snr} for the utilized \ac{snr} values.
$P\gets33$ for both users in Algorithm 2 so that a meaningful z-test
can be performed. 

\begin{figure}
\includegraphics{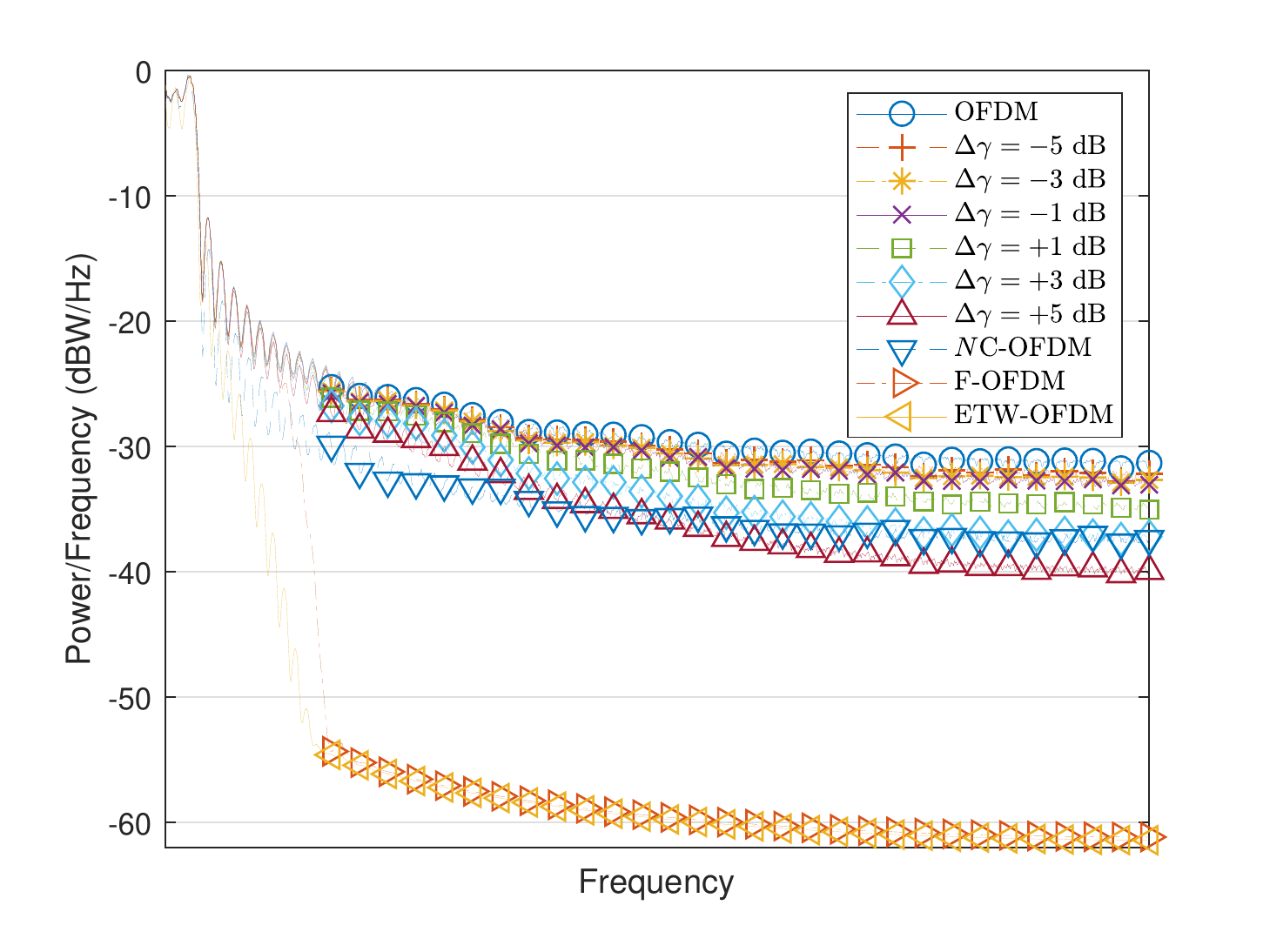}

\caption{\ac{oob} emission of investigated modulations.\label{fig:Fig4}}
\end{figure}

\begin{figure}
\includegraphics{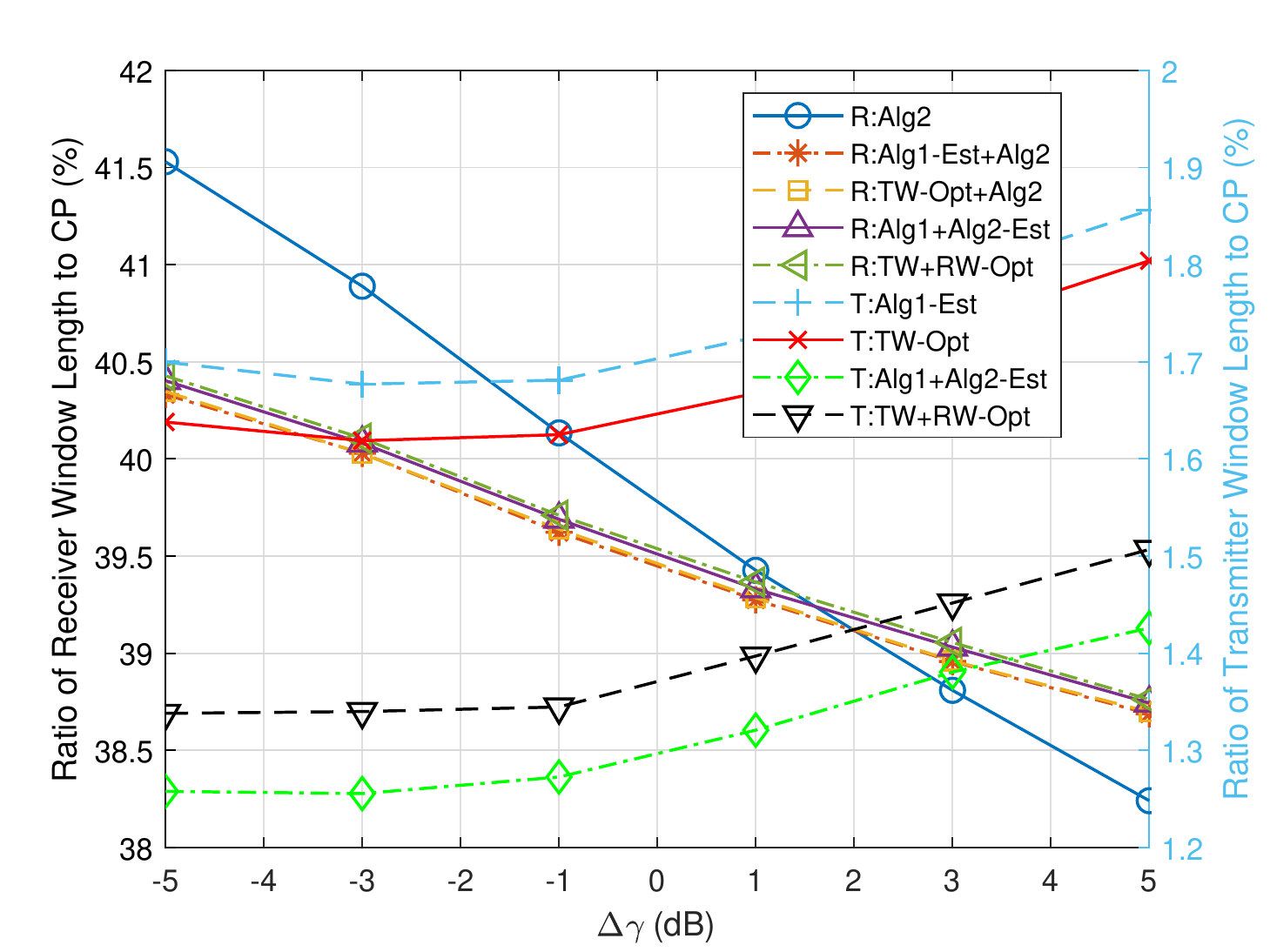}

\caption{$\mathbb{E}\left\{ \matr{R}_{u}/K_{u}\right\} $ and $\mathbb{E}\left\{ \matr{T}_{u}/K_{u}\right\} $
against the \ac{snr} difference between users.\label{fig:Fig5}}
\end{figure}

\begin{figure}
\includegraphics{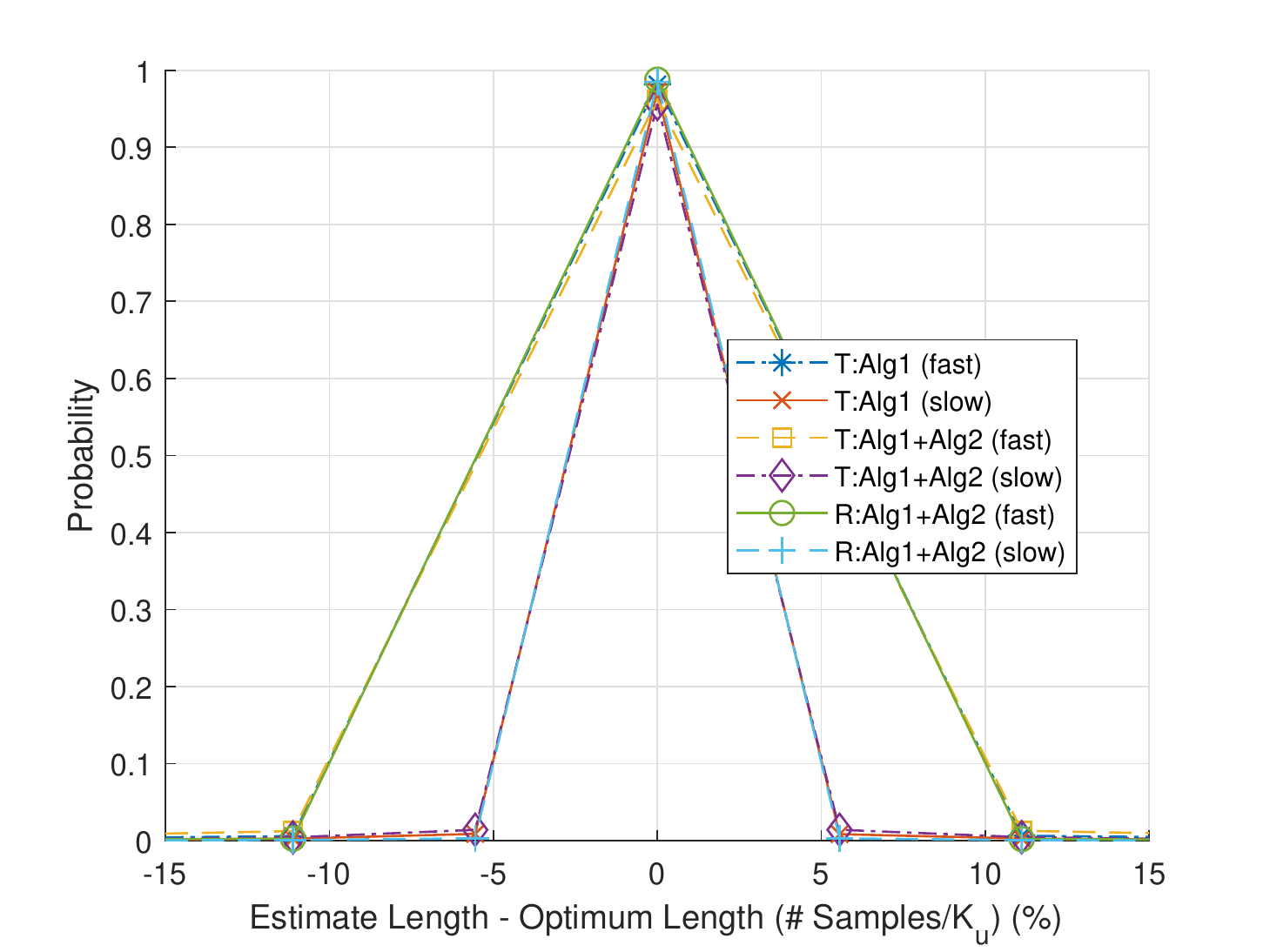}

\caption{Probability of the error between estimated and optimum window lengths
being equal to certain percentages of \ac{cp}.\label{fig:Fig6}}
\end{figure}

\begin{figure}
\includegraphics{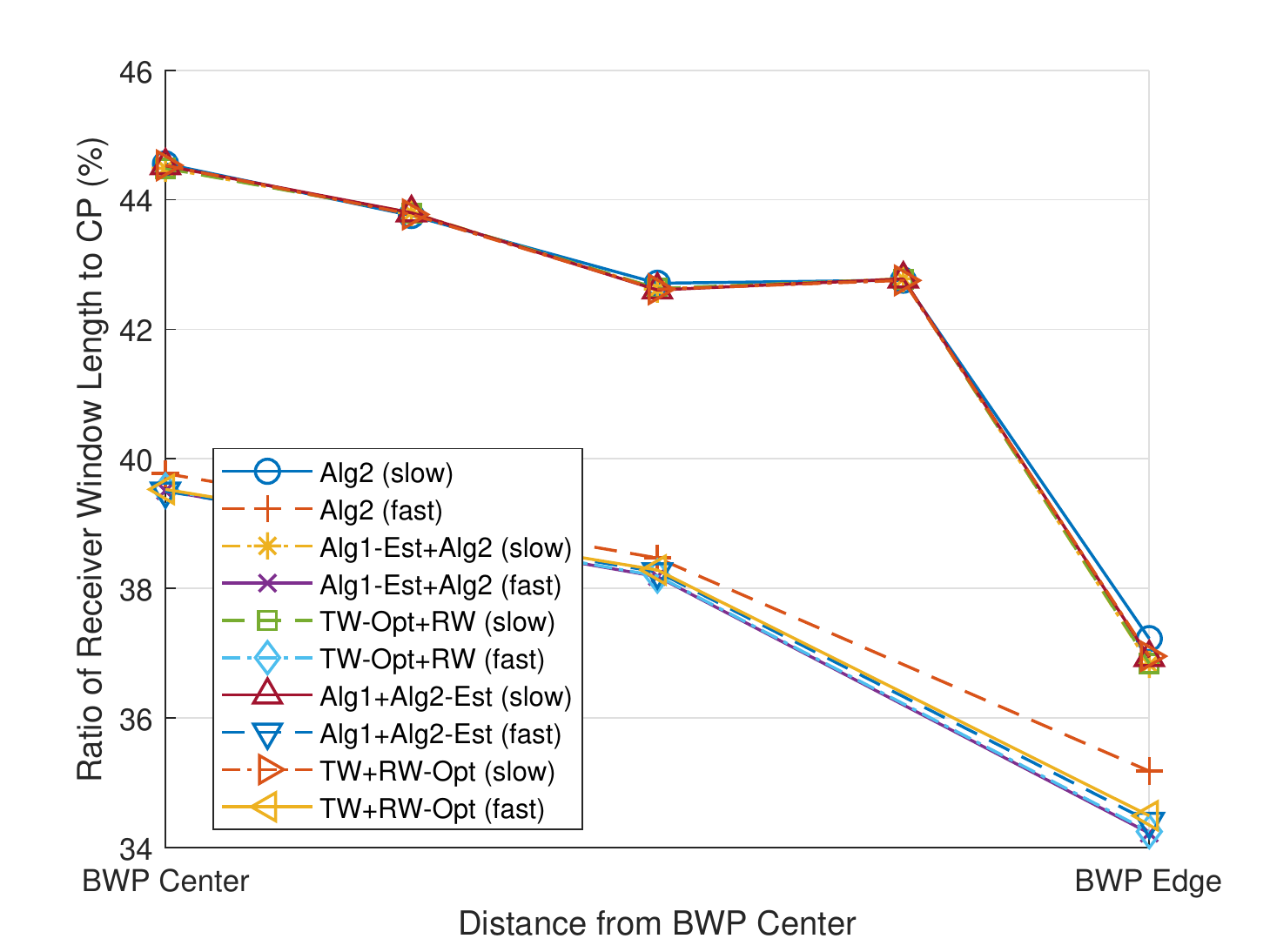}

\caption{Receiver windowing durations as a function of distance from center
of the consumed band.\label{fig:Fig7}}
\end{figure}

\begin{figure}
\includegraphics{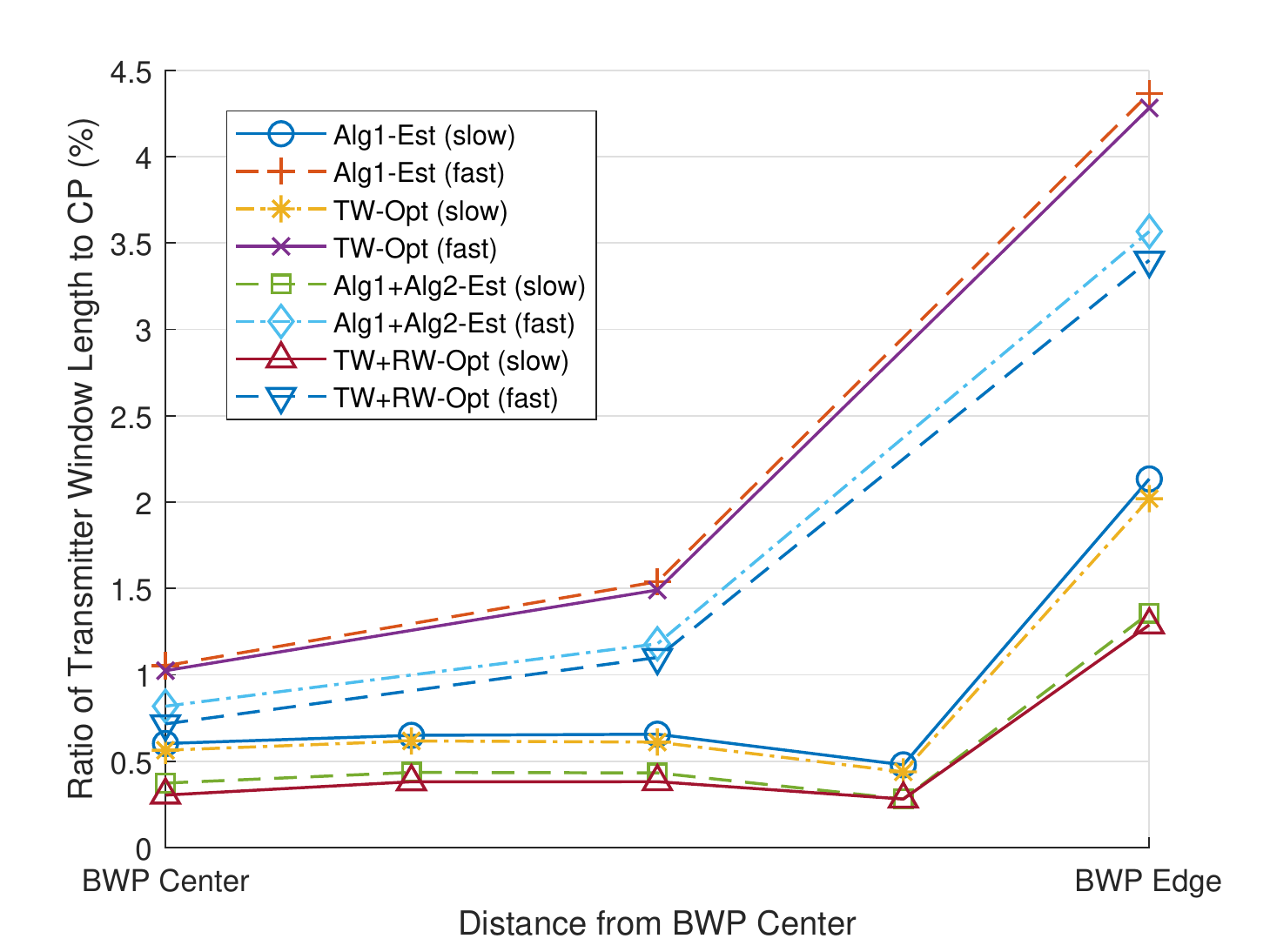}

\caption{Transmitter windowing durations as a function of distance from center
of the consumed band.\label{fig:Fig8}}
\end{figure}

\begin{figure}
\includegraphics{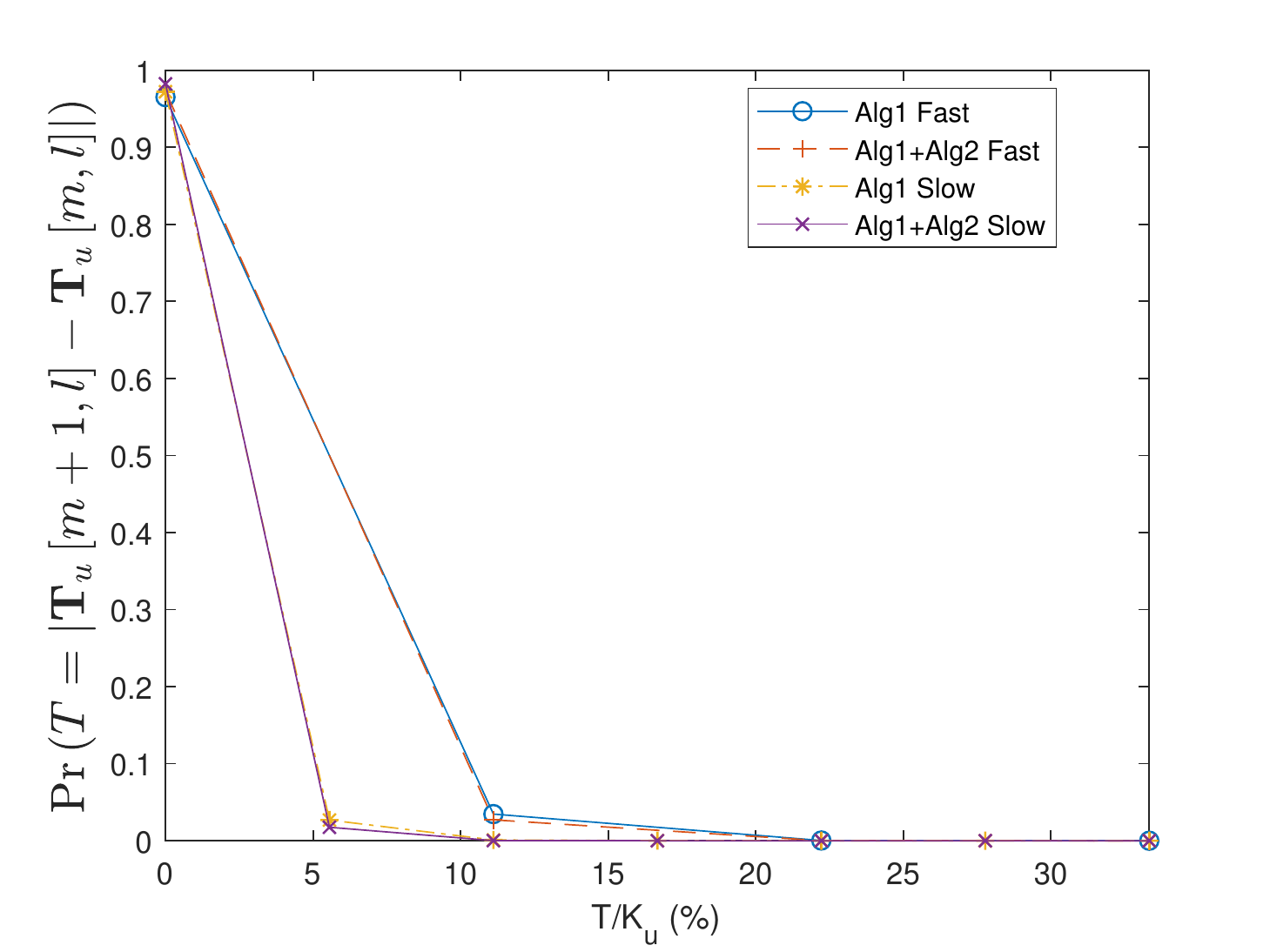}

\caption{Probability that transmitter window durations in adjacent subcarriers
differ by the given amount.\label{fig:Fig9}}
\end{figure}

\begin{figure}
\includegraphics{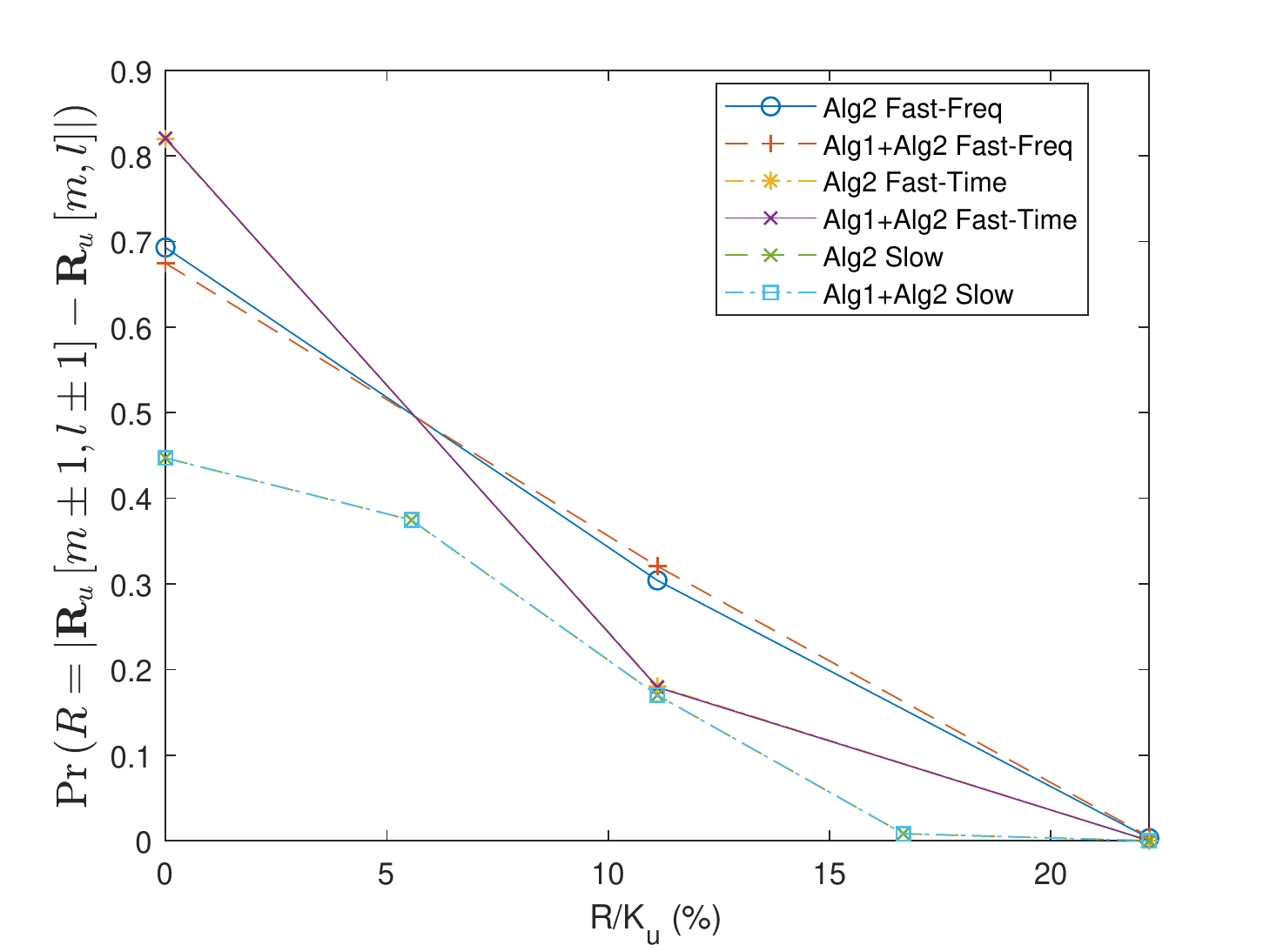}\caption{Probability that receiver window durations in adjacent \acp{re} differ
by the given amount.\label{fig:Fig10}}
\end{figure}

The \ac{oob} emission of investigated waveforms are depicted in \prettyref{fig:Fig4},
where the lines denoted with $\Delta\gamma$ is the results for Algorithm
1, for which the windowed user's average \ac{snr} is greater than
that of the victim by the provided value; and the sampling points
of the victim subcarriers are marked to distinguish between modulations
and to provide means to understand the unorthodox frequency localization
characteristics of \ac{fofdm}\citep{HuaweiFOFDM2016} and optimum
fixed extension transmitter windowing (ETW) algorithm\citep{guvenkaya2015awindowing}
to unfamiliar readers. Both \ac{fofdm} and ETW-OFDM have unmatched
interference performance in the victim's band, but \ac{fofdm} requires
the receiver to perform matched filtering, and ETW-OFDM requires an
extension that may disturb the standard frame structure, or reduced
throughput if the standard extensions are used in vehicular channels
as seen in \prettyref{tab:ThrSNR}. The interference performance of
\ac{ncofdm} at the edge subcarriers also outperforms all cases of
Algorithm 1, but Algorithm 1 takes over in the band center subcarriers
for high excess \ac{snr}. Furthermore, \ac{ncofdm} also requires
receiver-side operations, thus has no advantage over \ac{fofdm}.
It is seen that while Algorithm 1 has little advantage if the windowing
user has no excess \ac{snr}, the level of interference decreases
further as the window duration is able to increase when the user has
excess \ac{snr}. Although the proposed algorithm uses the same window
design used in ETW-OFDM, the fact that not all \acp{re} are windowed
prevents the same localization from surfacing. It should also be noted
that the gains are a significant function of channel responses of
both \acp{ue}, and the transmit \ac{oob} emission is unable to demonstrate
the gains clearly. 
The fair proportional network throughput, calculated similar to network
proportional network capacity using the geometric means of throughput
of each user, can be seen in \prettyref{tab:ThrSNR} for optimum fixed
extension transmitter and receiver windowed \ac{ofdm}\citep{guvenkaya2015awindowing},
\ac{ncofdm}, conventional \ac{cp}-\ac{ofdm}, adaptive transmitter
windowed with estimates obtained using Algorithm 1, adaptive transmitter
windowed with optimum durations, \ac{fofdm}, adaptive receiver windowed
using durations calculated using Algorithm 2, adaptive transmitter
and receiver windowed with transmitter windowing durations calculated
using Algorithm 1 without knowing receivers are applying Algorithm
2 followed by Algorithm 2 at the receivers, Algorithm 2 applied to
the signals that are adaptive transmitter windowed with optimum durations,
adaptive transmitter and receiver windowed with transmitter windowing
durations calculated using Algorithm 1 knowing receivers are applying
Algorithm 2 followed by Algorithm 2 at the receivers, and adaptive
transmitter and receiver windowed with durations optimized jointly.
The optimum values were obtained by maximizing the fair proportional
network throughput using an evolutionary integer genetic algorithm\citep{lin1992genetic}
to find the optimum inputs to Algorithms 1 and/or 2 under actual time-varying
channels. It can be seen that although previously proposed extended
windowing algorithms improve the \acp{ber}, increasing the effective
symbol duration by \textasciitilde 18\% erases the positive implications
on the throughput and reduces it. The artificial noise introduced
by the \ac{ncofdm} cannot be resolved at the receivers at these high
mobility conditions correctly yielding a decrease in actual throughput.
It can be seen that even the featured worst case results of the proposed
algorithms increase the throughput and improving algorithm outputs
by channel prediction promises further gains closer to optimum. While
\ac{fofdm} provides higher throughput compared to Algorithm 1 and
adaptive transmitter windowing, it requires that both ends of the
communication are aware of the filtering process and apply it\citep{turin_introduction_1960,abdoli_filtered_2015,HuaweiFOFDM2016}.
Although knowledge of such improves the throughput, the proposed algorithms
do not require the knowledge and action of the counterpart and this
is the strength of the proposed method compared to \ac{fofdm}.
\begin{table}
\caption{Fair proportional network throughput of tested modulations\label{tab:ThrSNR}}
\begin{tabular}{c|>{\centering}m{0.14\columnwidth}|>{\centering}m{0.15\columnwidth}}
Modulation & Throughput (Mbps) & Gain over CP-OFDM\tabularnewline
\hline 
\hline 
ETW-OFDM & 1.1949 & -15.832\%\tabularnewline
\hline 
ERW-OFDM & 1.1952 & -15.708\%\tabularnewline
\hline 
$N$C-OFDM & 1.3927 & -1.456\%\tabularnewline
\hline 
CP-OFDM & 1.3985 & -\tabularnewline
\hline 
Algorithm 1 & 1.3986 & +0.042\%\tabularnewline
\hline 
TW-OFDM (w/o Ext) & 1.3987 & +0.049\%\tabularnewline
\hline 
F-OFDM & 1.3988 & +0.057\%\tabularnewline
\hline 
Algorithm 2 & 1.3990 & +0.071\%\tabularnewline
\hline 
Algorithm 1 + Algorithm 2 (independent) & 1.3991 & +0.078\%\tabularnewline
\hline 
TW-OFDM + Algorithm 2 (independent) & 1.3992 & +0.085\%\tabularnewline
\hline 
Algorithm 1 + Algorithm 2 (joint) & 1.3996 & +0.114\%\tabularnewline
\hline 
TW-OFDM+ RW-OFDM (joint) & 1.3998 & +0.128\%\tabularnewline
\end{tabular}
\end{table}
To show the dependence of window durations on excess \ac{snr}, the
ratio of estimated and optimum expected window durations to the \ac{cp}
of the corresponding \acp{ue} as a function of the \ac{snr} difference
between the user in interest and the other user are demonstrated in
\prettyref{fig:Fig5}. The results are ordered as follows: Receiver
windowing durations of only Algorithm 2, Algorithm 2 applied to the
signals transmitted after applying Algorithm 1, Algorithm 2 applied
to transmitter windowed samples with the optimum duration if the \ac{gnb}
is unaware that receivers employ Algorithm 2, Algorithm 2 applied
to to the signals produced Algorithm 1 where \ac{gnb} knows both
receivers also employ Algorithm 2, and Algorithm 2 applied to transmitter
windowed samples with the optimum duration calculated knowing that
receiver will apply Algorithm 2; as well as transmitter windowing
durations estimated by Algorithm 1, optimum adaptive transmitter windowing
durations, transmitter windowing duration estimates provided by Algorithm
1 knowing that both receivers also employ Algorithm 2 and optimum
transmitter windowing durations calculated if both receivers also
employ Algorithm 2. A critical observation is that the transmitter
windowing durations, both estimated and actual optimum, increase as
the \ac{snr} of the user increases, whereas the receiver windowing
duration decreases. This proves the basic idea behind fair optimization
that the ones with excess \ac{snr} must focus on their impact on
others whereas the ones with lesser \ac{snr} must focus on the impact
they receive from others. It can also be seen that the optimum durations
for each side get shorter once the resources are jointly used, i.e.,
the \ac{gnb} knows that receivers utilize Algorithm 2.
\prettyref{fig:Fig6} shows the probability of the calculated window
duration being a certain amount away from the optimum duration, between
Algorithm 1 and optimum transmitter windowing durations, between Algorithm
1 calculated knowing that receivers utilize Algorithm 2 and optimum
transmitter window durations obtained when receivers employ Algorithm
2; and receiver windowing durations estimated at the transmitter during
calculation of Algorithm 2 and the values obtained at receivers. It
can be seen that the guess for both the transmitter and the receiver
windowing durations are more accurate for the slower user, proving
the dependence on mobility at estimates without channel tracking and
prediction. Furthermore, since receiver windowing durations only matter
for the RBs in interest as discussed before, receiver windowing durations
can be guessed with over 98\% probability without making an error.
The transmitter windowing estimates have more than 95\% probability
of being the same as optimum, while overestimating is slightly more
probable in the only Algorithm 1 case while underestimating is more
probable in the both algorithms utilized case.
\prettyref{fig:Fig7} and \prettyref{fig:Fig8}
show the amount of receiver and transmitter windowing applied at the
band centers and edges and checks the validity of \citep{Sahin2011}
where the window durations are labeled similar to that of \prettyref{fig:Fig5}.
It can be seen that the amount of transmitter windowing indeed increases
at the band edges, and furthermore it is more important that the faster
user with the larger subcarrier spacing and less spectral localization
to apply more transmitter windowing. This derives from the fact that
the \ac{psd} of signals with larger subcarrier spacing decay slower
than those with smaller subcarrier spacing, hence are more crucial
for the interference in the system. It can be seen that the receiver
windowing durations are higher at the band centers and higher for
the user with lower subcarrier spacing. This occurs partly due to
the window function design. The window functions are designed to minimize
the absorption outside the the band of interest, however as the pass-band
of the window gets smaller, the reduction performance decreases as
well\citep{guvenkaya2015awindowing}. Since the window pass-bands
are smaller on the edge subcarriers, the gain from reduced \ac{aci}
and \ac{ici} reduces whereas the performance reduction due to increased
\ac{isi} stays the same. This favors longer window durations at the
inner subcarriers where increasing window durations result in significant
\ac{ici} and \ac{aci} reduction. The gain from \ac{ici} reduction
becomes more prominent for the faster user which observes even higher
window lengths at inner subcarriers due to the increased \ac{ici}.
The gain from either type of windowing reduces for both users as windowing
at the counterparty is introduced to the systems, both by reduction
of forces driving windowing at a given side and also increase in \ac{isi}
occurring by applying windowing, as both users observe shorter windowing
durations on either side that is more uniformly distributed from band
centers to edges.
Before the average number of performed operations are provided for
presented Algorithms in their current forms and compared, spectrotemporal
statistics of window durations are provided to demonstrate that there
is room for further computational complexity reductions, which are
left for future works. Both experienced channel and amount of interference
are highly correlated in both dimensions, which in turn create correlated
window durations that can reduce complexity load. For example, \prettyref{fig:Fig9}
shows the probability that window durations calculated for adjacent
subcarriers differ by a given duration, as a function of \ac{cp}
length. It is seen that no more than $35\%$ \ac{cp} duration difference
occurred at any time. This suggests that if a subcarrier was calculated
to have a long window duration, checking brief window durations for
the adjacent subcarriers may be skipped at first and the search can
start from a higher value. Furthermore, \acp{re} may be grouped and
processed together. \prettyref{fig:Fig10} presents
the same results for Algorithm 2, showing that the differences are
even smaller in both time and frequency as the duration is determined
using the variance over a group of \acp{re} and the \ac{re} groups
of adjacent \ac{re} differ little. It is also worth noting that window
durations in adjacent \acp{re} of the faster user are more likely
to differ by longer durations than that of the slower user, which
depends on both increased subcarrier spacing and channel variations.
Finally, the computational load of the algorithms in their presented
forms is analyzed and compared with \ac{fofdm}. The filter lengths
are $N_{u}/2+1$ per \citep{HuaweiFOFDM2016}, and since filters consist
of complex values, the computational complexity of \ac{fofdm} is
$\left(N_{u}+K_{u}\right)L_{u}\left(3N_{u}/2+2\right)$ real additions
and $\left(N_{u}+K_{u}\right)L_{u}\left(2N_{u}+4\right)$ real multiplications
at the \ac{ue}, and these values summed over all users at the \ac{gnb}.
The computational complexities of the presented algorithms depend
on the window duration and side of each \ac{re}, of which values
have the probability distributions shown in \prettyref{fig:Fig11}.
Accordingly, the \ac{gnb} and \ac{ue} side computational complexities
of the algorithms are presented in \prettyref{tab:Computational-Complexities-of}.
As Algorithm 1 only runs at the \ac{gnb} and Algorithm 2 only runs
at the \ac{ue} without any operation requirements at the counterpart,
the counterpart complexities are 0 for both users. It is seen that
while the \ac{gnb} side complexity for Algorithm 1 is higher than
that of \ac{fofdm}, assumin that \acp{gnb} are not computationally
bounded, the transparency of Algorithm 1 still makes it a possible
candidate under heavy traffic. The computational complexity of Algorithm
2 is similar to that of \ac{fofdm} if the further computational complexity
reduction tricks described in the preceding paragraph are not employed,
and Algorithm 2 is also transparent to the transmitter. Another interesting
observation that can be made from \prettyref{fig:Fig11}
is that for Algorithm 2, under severe \ac{aci} conditions, longer
window durations may be beneficial, however since the window duration
is limited by \ac{cp} length, all those results manifest themselves
at the upper bound, creating a high probability peak at the longest
duration. Computational complexity can be further reduced if Algorithm
2 is modified to check the longest possibility before others, however
these highly implementation specific details are left for future work.
\begin{figure}
\includegraphics{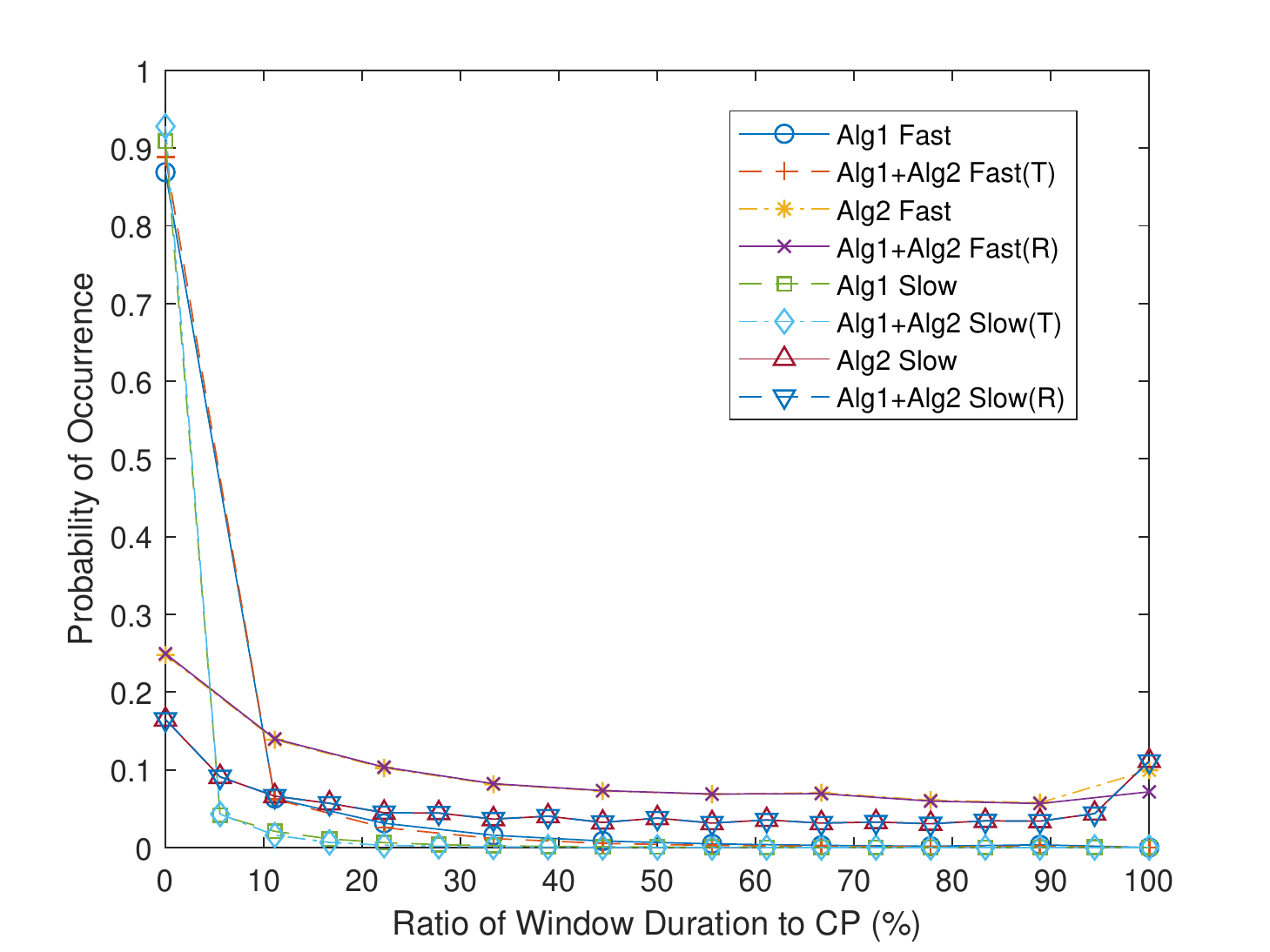}\caption{Probability of transmitter (T) and receiver (R) window durations occurring
in test scenarios.\label{fig:Fig11}}
\end{figure}
\begin{table}
\caption{Computational Complexities of \ac{fofdm} and Algorithms 1 and 2\label{tab:Computational-Complexities-of}}

\centering%
\begin{tabular}{c|c|c|>{\centering}m{0.15\columnwidth}|>{\centering}m{0.15\columnwidth}}
Algorithm & \ac{gnb} add & \ac{gnb} mult & \ac{ue} add & \ac{ue} mult\tabularnewline
\hline 
\hline 
\ac{fofdm} & 1907040 & 2551488 & 637872+ 1269168 & 854880+ 1696608\tabularnewline
\hline 
Alg. 1 & 5342353 & 5332295 & 0 & 0\tabularnewline
\hline 
Alg. 2 & 0 & 0 & 2088692+ 4166143 & 1054709+ 2236623\tabularnewline
\end{tabular}
\end{table}

\section{Conclusion\label{sec:Conclusion}}

In this work, we have demonstrated the concept of frame structure
compliant computationally efficient adaptive per-\ac{re} extensionless
transmitter windowing to maximize fair proportional beyond 5G network
capacity in the \ac{dl}, and universal per-\ac{re} receiver windowing
that requires no additional knowledge. Results demonstrate that gains
are possible from windowing without introducing extra extensions that
defy the frame structure if the side, \ac{re} and duration to apply
windowing is calculated carefully. The user with higher excess \ac{snr}
must apply longer transmitter windowing as they can resist the \ac{snr}
reduction, whereas the user with lower excess \ac{snr} must apply
longer receiver windowing. Users with higher subcarrier spacing and
higher mobility cause more interference in the system hence should
apply more transmitter windowing, whereas users with lower subcarrier
spacing must focus on receiver windowing. Optimum transmitter window
durations are longer at the edges whereas optimum receiver window
durations are longer at band centers. Emulating the multipath multiple
access channel allows the \ac{gnb} to estimate optimum transmitter
windowing durations prior to transmission with $95\%$ confidence.
Using the variance of received symbols allows the \acp{ue} to calculate
optimal receiver windowing durations without calculations requiring
further knowledge about the network and channel. While both algorithms
are presented for per-\ac{re} calculations, spectrotemporal correlation
of window durations allow reduced computational complexity implementations
than those described. Extensionless windowing at either side does
not require action and information transfer to the communication counterpart
and is fully compatible with previous and current generations, however
the knowledge of adaptive windowing applied at the counterpart allows
joint optimization that reveals higher gains.

\bibliographystyle{IEEEtran}
\bibliography{IEEEabrv,berker,3gpp_38-series}

\begin{IEEEbiography}[\includegraphics{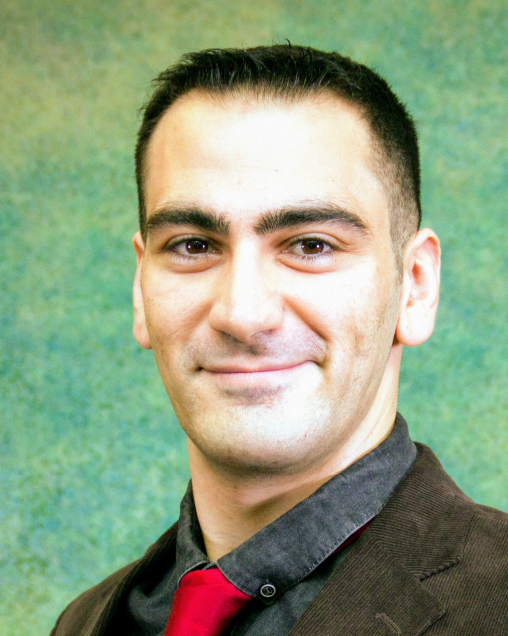}]{Berker Peköz}
(GS'15) received the B.S. degree in electrical and electronics engineering
from \acl{metu}, Ankara, Turkey in 2015, and the M.S.E.E. from \acl{usf},
Tampa, FL, USA in 2017.

He was a Co-op Intern at the Space Division of Turkish Aerospace Industries,
Inc., Ankara, Turkey in 2013, and a Summer Intern at the Laboratory
for High Performance DSP \& Network Computing Research, New Jersey
Institute of Technology, Newark, NJ, USA in 2014. He is currently
pursuing the Ph.D. at University of South Florida, Tampa, FL, USA.
His research is concerned with standard compliant waveform design
and processing.

Mr. Peköz is a member of Tau Beta Pi.
\end{IEEEbiography}

\begin{IEEEbiography}[\includegraphics{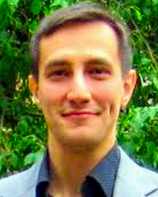}]{Selçuk Köse}
(S’10--M’12) received the B.S. degree in electrical and electronics
engineering from Bilkent University, Ankara, Turkey, in 2006, and
the M.S. and Ph.D. degrees in electrical engineering from the University
of Rochester, Rochester, NY, USA, in 2008 and 2012, respectively.

He was a Member of the VLSI Design Center at Scientific and Technological
Research Council (TÜBİTAK), Ankara, Turkey; the Central Technology
and Special Circuits Team of Enterprise Microprocessor Division at
Intel Corporation, Santa Clara, CA, USA; and the RF, Analog, and Sensor
Group at Freescale Semiconductor, Tempe, AZ, USA. He was an Assistant
Professor of Electrical Engineering at the University of South Florida,
Tampa, FL, USA. He is currently an Associate Professor of Electrical
Engineering at the University of Rochester, Rochester, NY, USA. His
current research interests include integrated voltage regulation,
3-D integration, hardware security, and green computing.

Dr. Köse was a recipient of the NSF CAREER Award, the Cisco Research
Award, the USF College of Engineering Outstanding Junior Researcher
Award, and the USF Outstanding Faculty Award. He is an Associate Editor
of the \emph{World Scientific Journal of Circuits, Systems, and Computers}
and the \emph{Elsevier Microelectronics Journal}. He has served on
the Technical Program and Organization Committees of various conferences.
\end{IEEEbiography}

\begin{IEEEbiography}[\includegraphics{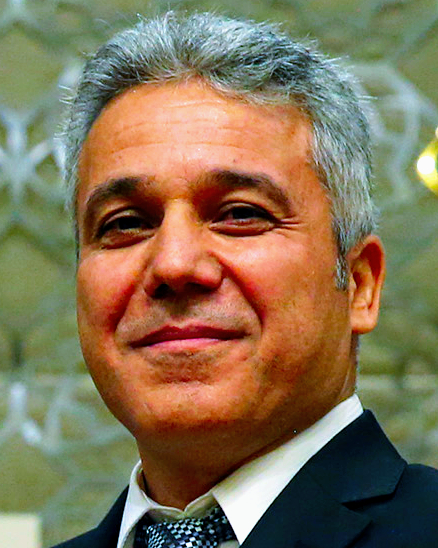}]{Hüseyin Arslan}
(S'95-M'98-SM'04-F'15) received the B.S. degree in electrical and
electronics engineering from \acl{metu}, Ankara, Turkey in 1992,
and the M.S. and Ph.D. degrees in electrical engineering from Southern
Methodist University, Dallas, TX, USA in 1994 and 1998, respectively.

From January 1998 to August 2002, he was with the research group of
Ericsson Inc., Charlotte, NC, USA, where he was involved with several
projects related to 2G and 3G wireless communication systems. He has
worked as part time consultant for various companies and institutions
including Anritsu Company (Morgan Hill, CA, USA), The Scientific and
Technological Research Council of Turkey (TÜBİTAK). He is currently
a Professor of Electrical Engineering at the University of South Florida,
Tampa, FL, USA, and the Dean of the College of Engineering and Natural
Sciences at the İstanbul Medipol University, İstanbul, Turkey.

Dr. Arslan’s research interests are related to advanced signal processing
techniques at the physical and medium access layers, with cross-layer
design for networking adaptivity and Quality of Service (QoS) control.
He has served as technical program committee chair, technical program
committee member, session and symposium organizer, and workshop chair
in several IEEE conferences. He is currently a member of the editorial
board for the \noun{IEEE Communications Surveys and Tutorials} and
the \emph{Sensors Journal}. He has also served as a member of the
editorial board for the \noun{IEEE Transactions on Communications},
the \noun{IEEE Transactions on Cognitive Communications and Networking
(TCCN)}, the \emph{Elsevier Physical Communication Journal}, the \emph{Hindawi
Journal of Electrical and Computer Engineering}, and \emph{Wiley Wireless
Communication and Mobile Computing Journal}.
\end{IEEEbiography}

\end{document}